\documentclass[twocolumn,nofootinbib,aps,prd,floats,floatfix,superscriptaddress,amsmath,amssymb,longbibliography,secnumarabic,preprintnumbers]{revtex4-1} %
\usepackage{graphicx}
\usepackage{dcolumn}
\usepackage[utf8]{inputenc}

\makeatletter
\def\thesection{\arabic{section}}%
\def\p@section{}%
\def\thesubsection{\thesection.\arabic{subsection}}%
\def\p@subsection{}%
\def\thesubsubsection{\thesubsection.\arabic{subsubsection}}%
\def\p@subsubsection{}%
\def\appendix{%
    \par
    \setcounter{section}\z@
    \setcounter{subsection}\z@
    \setcounter{subsubsection}\z@
    \def\thesubsection{\thesection.\arabic{subsection}}%
    \def\thesubsubsection{\thesubsection.\arabic{subsubsection}}%
    \def\p@subsection{}%
    \def\p@subsubsection{}%
    \@addtoreset{equation}{section}%
    \def\theequation@prefix{\thesection}%
    \addtocontents{toc}{\protect\appendix}%
    \@ifstar{%
    \def\thesection{\unskip}%
    \def\theequation@prefix{A.}%
    }{%
    \def\thesection{\Alph{section}}%
    }%
}%
\makeatother

\usepackage{bm}
\usepackage{hyperref}
\usepackage{xcolor}
\usepackage{float}
\usepackage{siunitx}
\usepackage[capitalise]{cleveref}
\usepackage[sort&compress]{natbib}
\usepackage{booktabs}

\begin{document}

\setlength{\abovedisplayskip}{5pt}
\setlength{\belowdisplayskip}{5pt}
\setlength{\abovedisplayshortskip}{5pt}
\setlength{\belowdisplayshortskip}{5pt}

\newcommand{\AdG}[1]{{\color{teal} \bf AdG: #1}}
\newcommand{\ims}[1]{{\color{blue} IMS: #1}}
\newcommand{\pbm}[1]{{\color{red} PBM: #1}}

\preprint{IFT-UAM/CSIC-26-128}

\title{Assessment of Super- and Hyper-Kamiokande Sensitivity to the DSNB}

\author{Pablo Blanco-Mas}
\email{pablo.blanco@ift.csic.es}
\affiliation{Instituto de Física Teórica UAM-CSIC, Calle Nicolás Cabrera 13–15, Universidad Autónoma de Madrid, 28049 Madrid, Spain}

\author{André de Gouvêa}
\email{degouvea@northwestern.edu}
\affiliation{Northwestern University, Department of Physics \& Astronomy, 2145 Sheridan Road, Evanston, IL 60208, USA}

\author{Iván Martínez-Soler}
\email{ivan.martinez@uam.es}
\affiliation{Departamento de F\'isica Te\'orica, Universidad Aut\'onoma de Madrid, Cantoblanco, 28049, Madrid, Spain}
\affiliation{Institute for Particle Physics Phenomenology, Department of Physics, Durham University, Durham DH1 3LE, UK}

\begin{abstract}

The detection of the Diffuse Supernova Neutrino Background (DSNB) is one of the main goals of current and next generation neutrino experiments. Based on results published by the Super-Kamiokande (SK) collaboration, we study the sensitivity of current and future water Cherenkov detectors to the DSNB. We reproduce current constraints and estimate the sensitivity to the DSNB of ten years of SK data and of a next-generation water-Cherenkov detector like Hyper-Kamiokande (HK). We find that, even with increased exposure, SK's sensitivity is relatively modest, while HK has the potential to deliver a statistically significant discovery of the DSNB in ten years. Along the way, we perform a detailed study of the relevant atmospheric neutrino backgrounds and argue that an accurate treatment of neutrino oscillations and propagation is essential when it comes to properly inferring the sensitivity to the DSNB and measuring its flux.

\end{abstract}

\maketitle

\section{Introduction}

Most of the energy released in a core-collapse supernova (SN) explosion is in the form of neutrinos. These carry invaluable information about the details of these extraordinary phenomena in their flavor, energy, and time distributions, assuming they can be detected and characterized here on Earth. However, the SN rate in our galaxy is rather small -- a few per century -- while the flux from individual SN explosions in nearby galaxies, which are much more abundant, is too small to be captured by existing and near-future neutrino detectors.   

There is, however, an intriguing alternative: the Diffuse Supernova Neutrino Background (DSNB)~\cite{Beacom_2010}. This is the isotropic, time-independent flux of the neutrinos emitted by all the past SN in the observable Universe. It is a steady source of neutrinos, and existing and near-future neutrino detectors are very well positioned to observe it, especially the water Cherenkov (WC) detectors Super-Kamiokande (SK), currently taking data, and Hyper-Kamiokande (HK), scheduled to start data taking in a few years.  

The DSNB neutrinos have energies around tens of MeV and come in all flavors and helicities~\cite{Beacom_2010}. At these energies, in WC detectors, the most relevant interaction channel is inverse beta decay (IBD), characterized by the coincident production of a positron and a neutron. Other processes are, in practice, not useful: at the relevant energies, neutral current interactions are difficult to detect in WC detectors, while the cross section for neutrino--electron scattering is highly suppressed relative to that for IBD. Consequently, SK and HK are most sensitive to the $\bar{\nu}_e$ component of the DSNB.

Backgrounds are formidable obstacles for the discovery of the DSNB. For antineutrino energies below 10~MeV, at SK and HK, IBD events associated with the DSNB are overwhelmed by those associated with the flux of electron antineutrinos from nearby nuclear reactors. This is an irreducible background. Also at WC detectors, the interaction of cosmic muons with oxygen nuclei produces spallation products, in particular $^{9}\mathrm{Li}$, which mimic the IBD signal for energies below $16~\mathrm{MeV}$~\cite{Super-Kamiokande:2025sxh, protocollaboration2018hyperkamiokandedesignreport}. This is the so-called spallation background. 

At higher energies, the dominant backgrounds to the DSNB are related to atmospheric neutrinos. Above roughly $35~\mathrm{MeV}$, the flux of atmospheric $\overline{\nu}_{e}$ exceeds all estimates for the $\bar{\nu}_e$ DSNB flux and also constitutes an irreducible background. Finally, for the energies of interest, charged- and neutral-current interactions of atmospheric neutrinos of all flavors and helicities with oxygen nuclei can also yield signals that mimic IBD, as demonstrated in~\cite{Zhou_2024}.

Here, we quantitatively estimate the sensitivity of SK and HK to the DSNB, assuming HK has the same neutron-tagging capabilities as SK loaded with gadolinium. We use the results reported by SK~\cite{Super-Kamiokande:2025sxh} to define the strength of the DSNB signal, to validate our calculations of the atmospheric neutrino backgrounds and, based on these, to assess the near- and intermediate-term potential of SK and HK to discover and characterize the DSNB. We also investigate the impact of neutrino oscillations on a potential DSNB detection. We show that an accurate treatment of atmospheric neutrino oscillations is required in order to properly infer the sensitivity to the DSNB and measure its properties, particularly for next-generation WC detectors such as HK.

The remainder of the paper is organized as follows. In Section~\ref{sec:dsnbflux}, we provide details on the DSNB flux. In Section~\ref{sec:atmflux}, we discuss the (low-energy) atmospheric neutrino flux. In Section~\ref{sec:osc}, we briefly introduce neutrino oscillations and discuss their impact on DSNB searches. Section~\ref{sec:events} details our simulation of DSNB-related events in SK and HK. Our results are presented and discussed in Section~\ref{sec:results}, followed by some concluding remarks in Section~\ref{sec:conclusions}. We include 
four appendices. In Appendix~\ref{sec:app-flux}, we give further details on the computation of the neutrino fluxes, and, in Appendix~\ref{sec:app-osc}, we provide further details on the neutrino-oscillation probabilities. In Appendix~\ref{combined}, we reproduce the preliminary results presented by Super-Kamiokande at Neutrino 2026. 
Finally, in Appendix~\ref{mixing}, we study the sensitivity to the oscillation parameters using atmospheric neutrinos with energies below several tens of MeV.

\section{The DSNB Flux}
\label{sec:dsnbflux}

The DSNB flux consists of all neutrinos emitted in core-collapse supernovae throughout the history of the Universe. We can parametrize this flux as \cite{de_Gouv_a_2020}
\begin{equation}\label{eq:DSNB}
\Phi_{\nu}(E)=\int_0^{z_{\rm max}}\frac{dz}{H(z)}\,\text{R}_{\rm CCSN}(z)\, F_\nu(E')\,,
\end{equation}
where $\text{R}_{\rm CCSN}$ is the star formation rate and $H(z)$ is the Hubble function that parametrizes the expansion of the Universe. The neutrino energy is red-shifted during propagation, such that $E = E'/(1+z)$. The neutrino spectrum is assumed to be thermal, approximated by a Fermi–Dirac distribution~\cite{Beacom_2010}:
\begin{equation}\label{eq:Nuspec}
F_\nu(E)=\frac{E_{\nu}^{\rm tot}}{6}\frac{120}{7\pi^4}\frac{E^2}{T_\nu^4}\frac{1}{e^{E/T_\nu}+1},
\end{equation}
where 
$E_{\nu}^{\rm tot}$ is the total energy emitted for that neutrino species.
This distribution depends on the neutrino temperature $T_{\nu}$ which in turn depends on the neutrino interaction cross section. We expect the hierarchy $T_{\nu_e} < T_{\overline{\nu}_e} < T_{\nu_x}$. Observations of the star formation rate indicate that most supernovae occur at redshifts $z \sim 1$--$2$.

The DSNB spectrum is, for the foreseeable future, only experimentally accessible inside a small energy window around $20$~MeV. Inside this energy range, we phenomenologically parameterize the flux with an exponential,
\begin{equation}\label{expflux}
    \Phi_\nu(E)=\Phi_0\, e^{-(E-E_0)/T_{\rm eff}},
\end{equation}
\noindent where $\Phi_{0}$ is the flux at a reference energy $E_{0}$ and $T_{\rm eff}$ is an effective temperature. This constitutes a simplified DSNB flux model, but it is nonetheless a useful description of the spectral shape in the energy range accessible to experiments. 

For a given species, the effective temperature $T_{\rm eff}$ can be approximately related to the neutrino emission temperature by
\begin{equation}
\frac{1}{T_{\rm eff}} = \frac{1+z_{\rm eff}}{T_\nu}\frac{1}{1+e^{-E_{0}/T_{\nu}}}-\frac{2}{E_{0}}.
\label{Teff}
\end{equation}
where $z_{\rm eff}$ denotes the effective redshift at which most of the neutrinos contributing to the observed flux are produced. For neutrino temperatures relevant to core-collapse supernovae, $T_\nu \sim 6$~MeV, and, for a reference energy $E_0 \sim 25$~MeV, $E_0/T_\nu \gg 1$. In this limit, Eq.~\eqref{Teff} simplifies to
\begin{equation}\label{eq:teff}
\frac{1}{T_{\rm eff}} = \frac{1+z_{\rm eff}}{T_\nu},
\end{equation}
so $T_{\rm eff}\simeq T_\nu$ for typical values of $z_{\rm eff}$ and $E_0$.

The effective temperature controls how the DSNB spectrum decreases with energy and thus allows one to investigate whether the emitted neutrinos follow a thermal distribution. A measurement of $T_{\rm eff}$ is also relevant for assessing the impact of black hole formation on the neutrino spectrum or the presence of distortions in the spectral shape associated with the cosmic star formation history~\cite{Keil:2002in,Kresse:2020nto,Ashida:2023heb,Nakazato:2024gem,Charisse:2025goq}.

\section{Atmospheric Neutrino Flux}
\label{sec:atmflux}

The atmospheric neutrino flux at tens of MeV constitutes one of the dominant and irreducible backgrounds for DSNB searches. An accurate determination of its normalization, flavor composition, and spectral shape is therefore essential for the extraction of a potential DSNB signal. A quantitative description of this flux requires detailed simulations of cosmic-ray interactions and particle transport.

Atmospheric neutrinos are produced by the interactions of cosmic rays with nuclei in the Earth's atmosphere. These translate into a constant flux of mesons -- primarily pions and kaons -- which decay into muons and muon neutrinos. At energies roughly below one GeV, muons are not highly boosted and decay before reaching the Earth's surface into electrons, electron neutrinos and muon neutrinos, also contributing to the atmospheric neutrino flux. This process results in a spectrum of both $\nu_{\mu}$ and $\nu_{e}$, plus their antiparticles, that spans a wide range of energies, from tens of MeV to tens of TeV~\cite{Gaisser:2002jj}.

Atmospheric muons with energies above one GeV typically reach the surface of the Earth before decaying. As they propagate through the Earth, they lose energy via ionization and eventually decay at rest, contributing to the low-energy (tens of MeV) atmospheric neutrino flux~\cite{Battistoni:2002ew}.

We computed the flux of atmospheric neutrinos at the SK and HK sites using FLUKA, a high-performance, general-purpose Monte Carlo simulation package used to model the interaction and transport of particles and nuclei in matter \cite{FLUKAwebsite,Ahdida:2022fluka,Battistoni:2015fluka,Hugo:2024fluka,Donadon:2024flair3,Roesler:2001dpmjet,Fedynitch:2015thesis,Battistoni_2011}. We simulated the transport and interactions of the cosmic rays and the transport and decay of the daughter particles into neutrinos in the Earth's atmosphere, assuming a dipole magnetic field for the planet. To simulate SK's overburden, we added a final one-km layer of silicon to the simulation and included the low-energy contribution to the flux coming from muon decay at rest. Throughout, we assumed the flux to be up--down symmetric.

\begin{figure}
    \centering
    \includegraphics[width=\columnwidth]{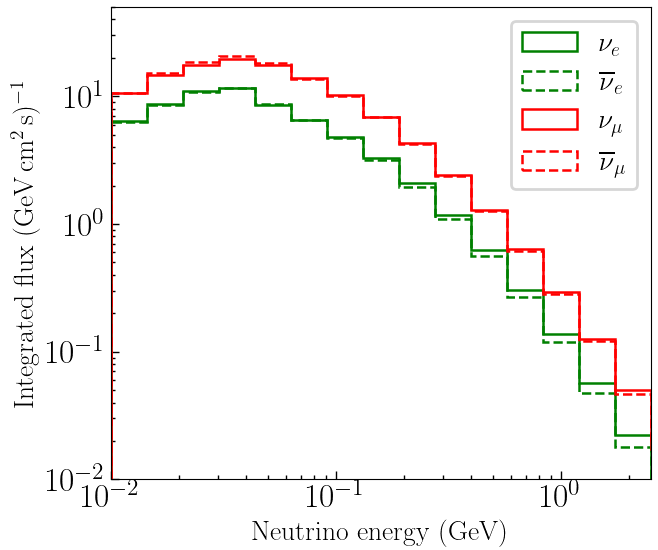}
    \caption{Unoscillated atmospheric neutrino and antineutrino fluxes as a function of the neutrino energy. These were computed with Fluka.}
    \label{fig:atm_flx}
\end{figure}

Our result for the atmospheric neutrino fluxes as a function of the neutrino energy is depicted in Figure~\ref{fig:atm_flx}. They are consistent with existing atmospheric flux calculations, such as those by Honda et al.~\cite{Honda_2011}, within the expected theoretical uncertainties. See Appendix~\ref{sec:app-flux} for a comparison with other existing calculations.

\section{Flavor oscillation}
\label{sec:osc}

The neutrino fluxes depicted in Fig.~\ref{fig:atm_flx} do not take into account neutrino masses and lepton mixing. SK has studied in detail the impact of flavor oscillations on the atmospheric neutrino flux~\cite{Super-Kamiokande:1998kpq}. In all recent analyses, oscillations are implemented within a full three-flavor framework as a function of energy and zenith angle. While this treatment naturally extends to the low-energy range relevant for DSNB searches, the impact of neutrino oscillations in this regime has not been explicitly discussed in experimental analyses.

\begin{figure}
    \centering
    \includegraphics[width=\columnwidth]{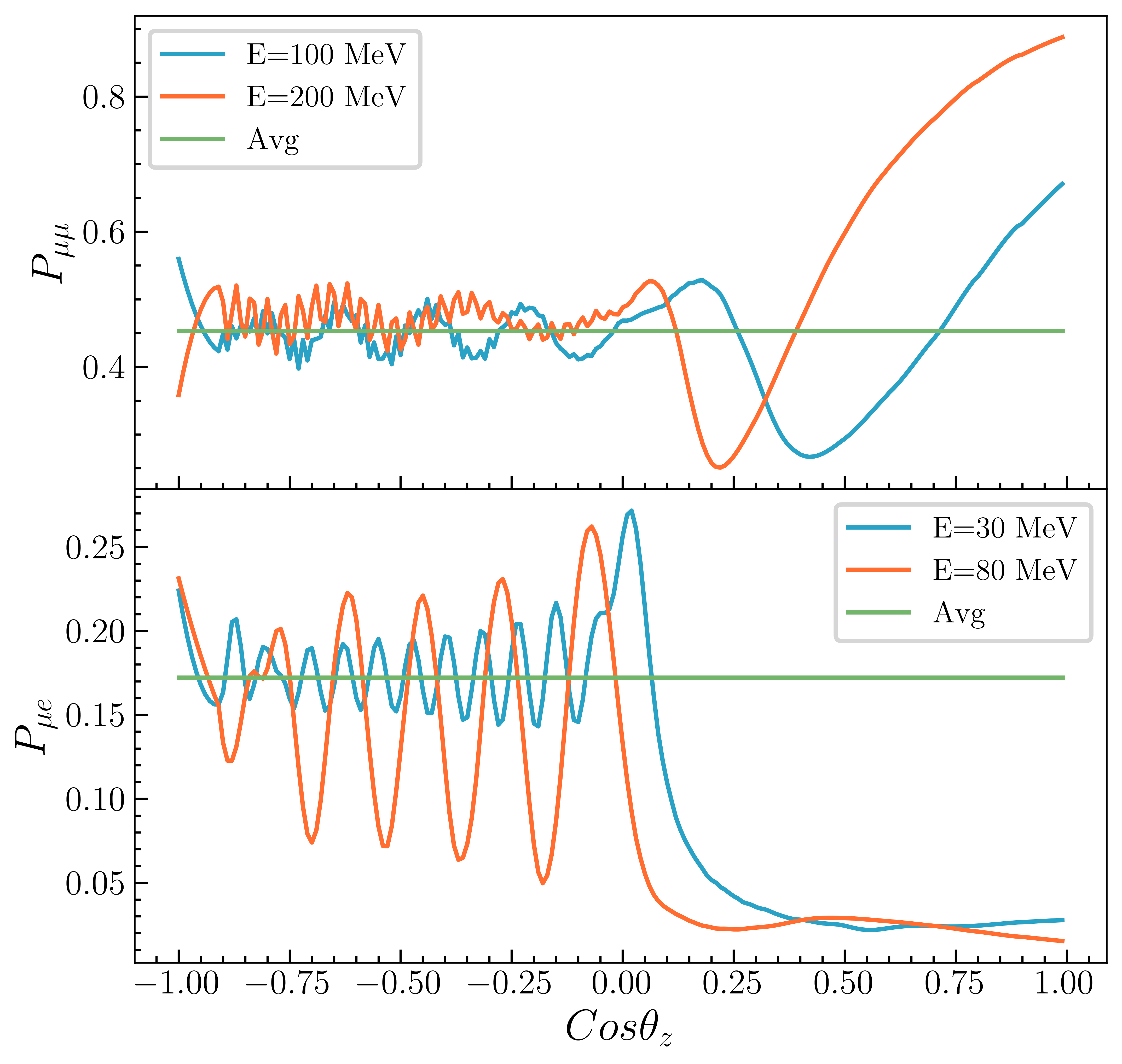}
    \caption{Muon neutrino oscillations as a function of the cosine of the zenith angle. The oscillation probabilities for muon disappearance (top) and electron appearance (bottom) are shown for two energies. In each case, we compare these results with the corresponding averaged oscillation probability (Avg).}
    \label{fig:Osc}
\end{figure}

Figure~\ref{fig:Osc} depicts the $\nu_{\mu}$ survival probability and the $\nu_{\mu}\to\nu_e$ oscillation probability as a function of the zenith angle for two representative energies,\footnote{See Appendix~\ref{sec:app-osc} for the oscillation probabilities for all baselines and energies between $10$~MeV and $1$~GeV.} where we have integrated over the neutrino production point. Unless otherwise stated, we use the best-fit values of the oscillation parameters from NuFit~\cite{Esteban:2024eli} throughout this work, and assume the neutrino mass ordering is normal.
It is clear that oscillation effects are non-trivial and directly impact the atmospheric background relevant for DSNB searches.

For energies below $100$~MeV and for trajectories crossing the Earth ($\cos\theta_z<0$), the oscillation length associated with $\Delta m^2_{21}$ is shorter than the distance traveled by the neutrinos. Integration over these trajectories results in an averaged-out oscillation probability, as illustrated in Figure~\ref{fig:Osc}. In contrast, down-going neutrinos ($\cos\theta_z>0$) propagate only through the atmosphere, with baselines ranging from tens of kilometers for vertical trajectories to hundreds of kilometers for horizontal ones, depending on the production height. In this regime, oscillations are not fully averaged out. Instead, for energies below $100$~MeV, the oscillation length associated with $\Delta m^2_{21}$ is comparable to the baselines corresponding to horizontal trajectories whereas for more vertical directions the propagation distance is comparable to the oscillation length driven by $\Delta m^2_{31}$. This translates into energy-dependent effects that are only partially washed out by the distribution of the neutrino production points. 

\section{Event Simulation} 
\label{sec:events}

\begin{figure*}[ht]
    \centering
    \includegraphics[width=0.7\textwidth]{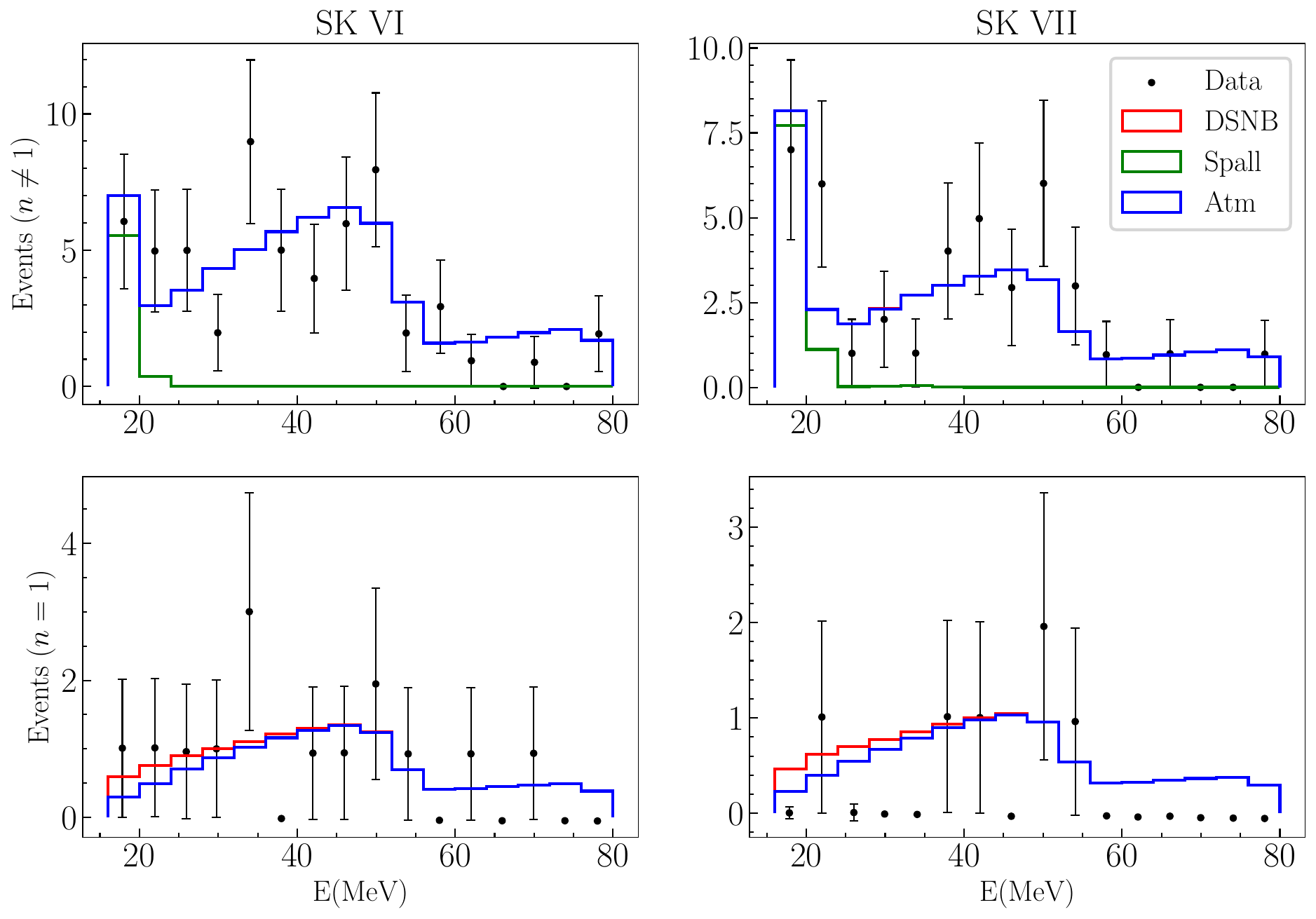}
    \caption{Event distributions for SK-VI and SK-VII as a function of the reconstructed lepton energy. The bottom (top) panels show the cases with one (zero or more than one) identified neutrons. We have separated the contributions from the spallation background (green), the atmospheric background (blue), and the DSNB prediction (red).}
    \label{fig:Events}
\end{figure*}

The expected number of observed events is proportional to the size of the detector, the amount of time dedicated to collecting the data, and the convolution of the neutrino flux, the interaction cross section, and the detector efficiency. The number of charged leptons (electrons and positrons) in the $i$-th reconstructed charged-lepton
energy bin, assuming a charged-current (CC) neutrino scattering event, is
\begin{equation}
    N_{i} = T\, N_{\text{targ}} \int dE_{\nu}\, dE_{l}\, dE_{r}\; \phi(E_{\nu})\, \frac{d\sigma}{dE_{l}}\, \varepsilon(E_{l},E_r)\, \mathrm{Eff}_{i} \, ,
\label{eq:N}
\end{equation}
where $E_{\nu}$ is the neutrino energy, $E_{l}$ is the energy of the outgoing charged lepton, and $E_{r}$ is the reconstructed charged-lepton energy, restricted to the $i$-th energy bin. $N_{\text{targ}}$ denotes the number of target nucleons, and $T$ is the exposure time; we consider a detector with a fiducial volume of $22.5$~kton.~\footnote{When simulating HK data, we assume a volume eight times larger.} The detector response is modeled through the smearing function $\varepsilon(E_{l},E_{r})$, which accounts for the finite energy resolution, and $\mathrm{Eff}_i$, the bin-dependent detection efficiency. We assume a Gaussian uncertainty in the lepton energy reconstruction of $6\%$ for $E_{\nu}=$30~MeV, scaling with energy as $\sigma/\text{MeV} = 0.35\sqrt{E/\text{MeV}}$~\cite{Super-Kamiokande:2023jbt}. The detector efficiency ($\mathrm{Eff}_{i}$), taken from~\cite{Super-Kamiokande:2025sxh}, depends on the data-collection phase -- here we consider the SK-VI and SK-VII phases -- as well as the number of reconstructed neutrons. In this work, we adopt the efficiency corresponding to the neural-network-based neutron-tagging method in\cite{Super-Kamiokande:2025sxh}. 

In Eq.~\eqref{eq:N}, $d\sigma/dE_l$ is the differential cross section as a function of the final-state charged-lepton energy for the relevant neutrino scattering processes. DSNB neutrinos are predominantly detected by IBD, $\bar{\nu}_e + p \to n + e^+$, with subleading contributions from other interaction channels. In addition to IBD, we include neutrino scattering on oxygen nuclei and electrons. These contribute to both the DSNB signal and, in some cases, to the background.

We simulate neutrino interactions in water using the \texttt{NuWro} event generator~\cite{Golan:2012rfa,Golan:2012wx}, which provides a detailed description of the interaction kinematics and final-state particle production. The simulation includes CC interactions for all neutrino flavors, as well as final-state interactions (FSI) within the nucleus. We validated the total cross sections obtained with \texttt{NuWro} against theoretical calculations for IBD and neutrino--oxygen scattering~\cite{Vogel:1999zy,Strumia:2003zx,Nakazato:2018xkv}, finding good agreement among them~\cite{Vogel:1999zy,Nakazato:2018xkv}. 

The scattering of neutrinos and antineutrinos off oxygen nuclei can break up the nucleus, leading to the emission of a neutron~\cite{Ankowski:2013gha,Ankowski:2011ei,Zhou_2024}. In the case of neutrino scattering, FSI, the re-scattering of nucleons inside the nucleus, can also result in neutron emission~\cite{Golan:2012wx}, thereby also contributing to the production of free neutrons. Additionally, although less relevant at these energies, the so-called 2p2h interactions, in which neutrinos scatter off correlated pairs of nucleons via meson-exchange currents, can also lead to neutron emission.

The CC scattering of $\nu_{\mu}$ and $\bar{\nu}_{\mu}$ occurs only at center-of-mass energies above the muon mass. When the daughter muon is not sufficiently energetic, it will not emit Cherenkov radiation. This occurs for energies below
\begin{equation}\label{eq:cherenkov}
E^{\rm th}_\mu = \frac{m_\mu c^2}{\sqrt{1-\frac{1}{n^2}}} ,
\end{equation}
where $n$ is the refractive index. In this regime, the untagged muon decays into an electron or positron, producing a signal in the detector similar to that of IBD. These leptons have energies up to $\sim 50$~MeV (see Fig.~\ref{fig:Events}). This contribution dominates the background at low energies; we simulate it by including muon decay in the events generated with \texttt{NuWro}.

Figure~\ref{fig:Events} depicts the simulated number of events for the different reconstructed outgoing positron and electron energy bins. 
We simulate data for SK-VI and SK-VII, corresponding to exposures of 552 and 404 days, respectively~\cite{Super-Kamiokande:2025sxh}. The events are distributed in 16 bins of $4$~MeV width, starting from $16$~MeV.
We classify events depending on the number of outgoing neutrons. In particular, we are interested in distinguishing events with one neutron from those with either zero or more than one neutron, as this information plays a crucial role in separating signal from background. The green line in Fig.~\ref{fig:Events} represents the expected event distribution from spallation processes, which we discuss momentarily, while the blue line represents the expected event distribution from all backgrounds included in the analysis, namely spallation and atmospheric neutrinos. The red line represents the contribution from the DSNB, for which we assume the Horiuchi et al.~(2009) model with an effective temperature of $6$~MeV~\cite{Horiuchi:2008jz}.  Figure~\ref{fig:Events} also includes the data from SK, reported in \cite{Super-Kamiokande:2025sxh}.

Cosmic muons can induce spallation processes that break up oxygen nuclei~\cite{Li:2014sea,Li:2015kpa,Super-Kamiokande:2021snn}. Although most spallation events produce a single electron, some of the resulting isotopes, such as $^{9}\text{Li}$, can emit a neutron during their decay. The relatively long lifetimes of these isotopes make them a non-negligible background for the DSNB. These backgrounds are particularly relevant at energies below $20$~MeV and are included in Fig~\ref{fig:Events} (green line).

Finally, for all neutrino flavors, neutral-current scattering off oxygen also contributes. In such events, a nucleon is ejected and the residual daughter nucleus promptly de-excites via gamma-ray emission. The de-excitation pattern depends on the oxygen shell from which the nucleon is removed. In some cases, the excitation energy exceeds $10$~MeV, leading to event topologies similar to those of IBD. Since this background is more relevant for energies below $16$~MeV, it is not included in our analysis.

\section{Results}
\label{sec:results}

This section contains our main results. We first validate our simulation framework by reproducing the current sensitivity of SK to the DSNB. We then use this validated setup to project the sensitivity of future experiments, including SK after ten years of exposure and ten years of HK data. Finally, we quantify the impact of neutrino oscillations on the atmospheric background and on the inferred DSNB sensitivity.

\subsection{Sensitivity to the DSNB}\label{sec:results_overall}

 We reproduce the sensitivity of SK to the DSNB using the published data. We focus on the measurements from phases SK-VI and SK-VII~\cite{Super-Kamiokande:2025sxh}.\footnote{We briefly discuss the results presented at the Neutrino 2026 Conference in Appendix~\ref{combined}} For the calculation of the expected number of events, we use the efficiencies provided by the collaboration. In the analysis, we include event samples with one neutron ($n=1$) as well as those with zero or more than one neutron ($n \neq 1$). Further details on the simulation of neutrino scattering and event reconstruction can be found in Section~\ref{sec:events}.

In order to test the sensitivity to the DSNB we defined a binned Poissonian $\chi^2$
 \begin{equation}
     \chi^2_{0,exp} = \rm 2\sum_{i,j} T^{i,j}_{exp} - O^{i,j}_{exp} + O^{i,j}_{exp} ln\left(\frac{O^{i,j}_{exp}}{T^{i,j}_{exp}}\right),
 \end{equation}
where $\rm T^{i,j}$ and $\rm O^{i,j}$ refer to predicted and observed events respectively, the $i,j$ indices run over all energy bins considered and over both $n=1$ and $n\neq 1$, and the ``exp" label goes over SK-VI and SK-VII. For the combined analysis we define a total combined $\chi^2_{0,tot}$ as
\begin{equation}
    \chi^2_{0,tot}=\rm \chi^2_{0,SK-VI}+\chi^2_{0,SK-VII}.
\end{equation}
We include uncertainties on the atmospheric neutrino flux that account for the overall normalization, the flavor composition, and the $\nu/\overline{\nu}$ ratio. The atmospheric neutrino flux has been measured by Super-Kamiokande over more than 20 years, primarily at higher energies. 
The flux below 100~MeV, however, is the product of the same underlying mechanisms -- pion and muon decay -- so we expect these uncertainties to be similar. Following the Super-Kamiokande analysis~\cite{Super-Kamiokande:2017yvm}, we adopt a $25\%$ uncertainty in the overall normalization and a $2\%$ uncertainty in both the flavor composition and the neutrino-to-antineutrino ratio.

We take the cosmic muon background from the Super-Kamiokande prediction~\cite{Super-Kamiokande:2025sxh}. As this background primarily affects the first energy bin, we include a free normalization for this component. Additionally, we allow for an independent normalization between the two phases (SK-VI and SK-VII), as well as between the two samples ($n=1$ and $n \neq 1$).

\begin{figure}
    \centering
    \includegraphics[width=\columnwidth]{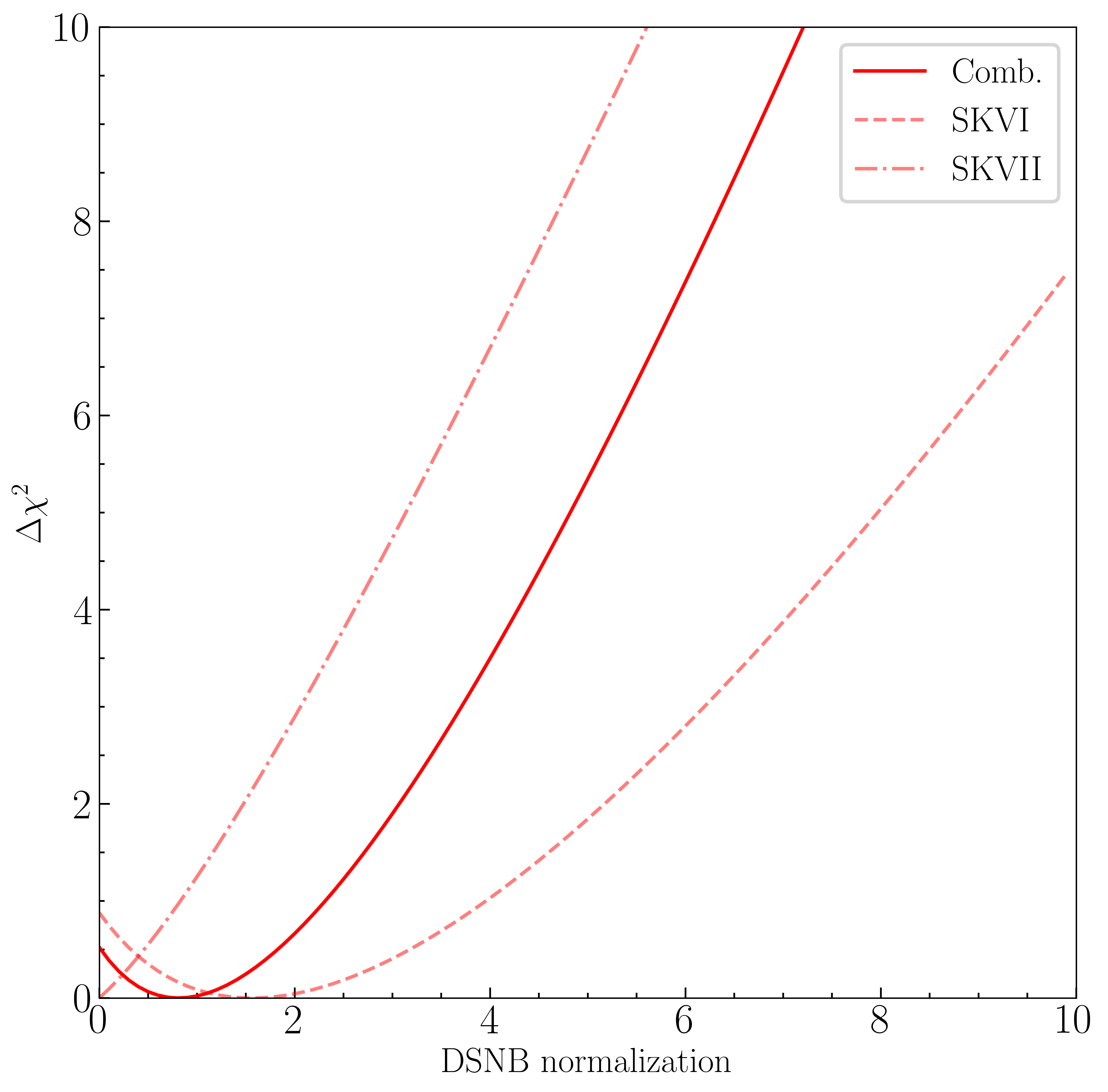}
    \caption{SK-VI, SK-VII, and combined analyses including the $n=1$ and $n \neq 1$ samples. The DSNB normalization is defined relative to the Horiuchi et al.~(2009) model with an effective temperature of $6$~MeV~\cite{Horiuchi:2008jz}. Our results agree with those in \cite{Super-Kamiokande:2025sxh}.}
    \label{fig:fit_actual}
\end{figure}

Assuming that the DSNB follows a thermal spectrum with a  temperature of $6$~MeV~\cite{Horiuchi:2008jz}, we perform a fit to the SK-VI and SK-VII data, shown in Fig.~\ref{fig:fit_actual}. The DSNB normalization is defined relative to the Horiuchi et al.~(2009) model with an effective temperature of $6$~MeV~\cite{Horiuchi:2008jz}. We find that these SK data do not translate into significant evidence for the DSNB. On the other hand, flux normalizations larger than 7 (relative to that of the Horiuchi et al.~(2009) model with an effective temperature of $6$~MeV~\cite{Horiuchi:2008jz}) are excluded at the three-sigma level ($\Delta\chi^2 \ge 9$) This corresponds to $10\,\nu/\text{cm}^2\,\text{s}$ with energies above $16$~MeV. The fit indicates a mild preference for a non-zero DSNB flux, around $1\,\nu/\text{cm}^2\,\text{s}$ with energy above 16~MeV, but with low statistical significance ($\Delta\chi^2 \simeq 0.5$ for the no-DSNB hypothesis).

We explored how the sensitivity changes when the systematics associated with the atmospheric neutrino flux change. In particular, we considered several scenarios, ranging from no prior knowledge (treated as a free parameter) to perfect knowledge (fixed parameter) of different aspects of the atmospheric neutrinos. We find that the uncertainty in the $\nu/\overline{\nu}$ ratio has only a minor impact on the results. For the uncertainty in the initial flavor composition, allowing it to float unconstrained leads to a preference for a lower DSNB rate; however, the best fit corresponds to a suppression of the electron component of the atmospheric flux by a factor of $\sim 2$. Regarding the overall normalization of the atmospheric flux, our benchmark flux normalization stays close to Honda's prediction. When this parameter is left free, we find a preference for a higher DSNB rate, although associated with a large suppression of the atmospheric neutrino flux normalization (approximately $30\%$).

\begin{figure}
    \centering
    \includegraphics[width=\columnwidth]{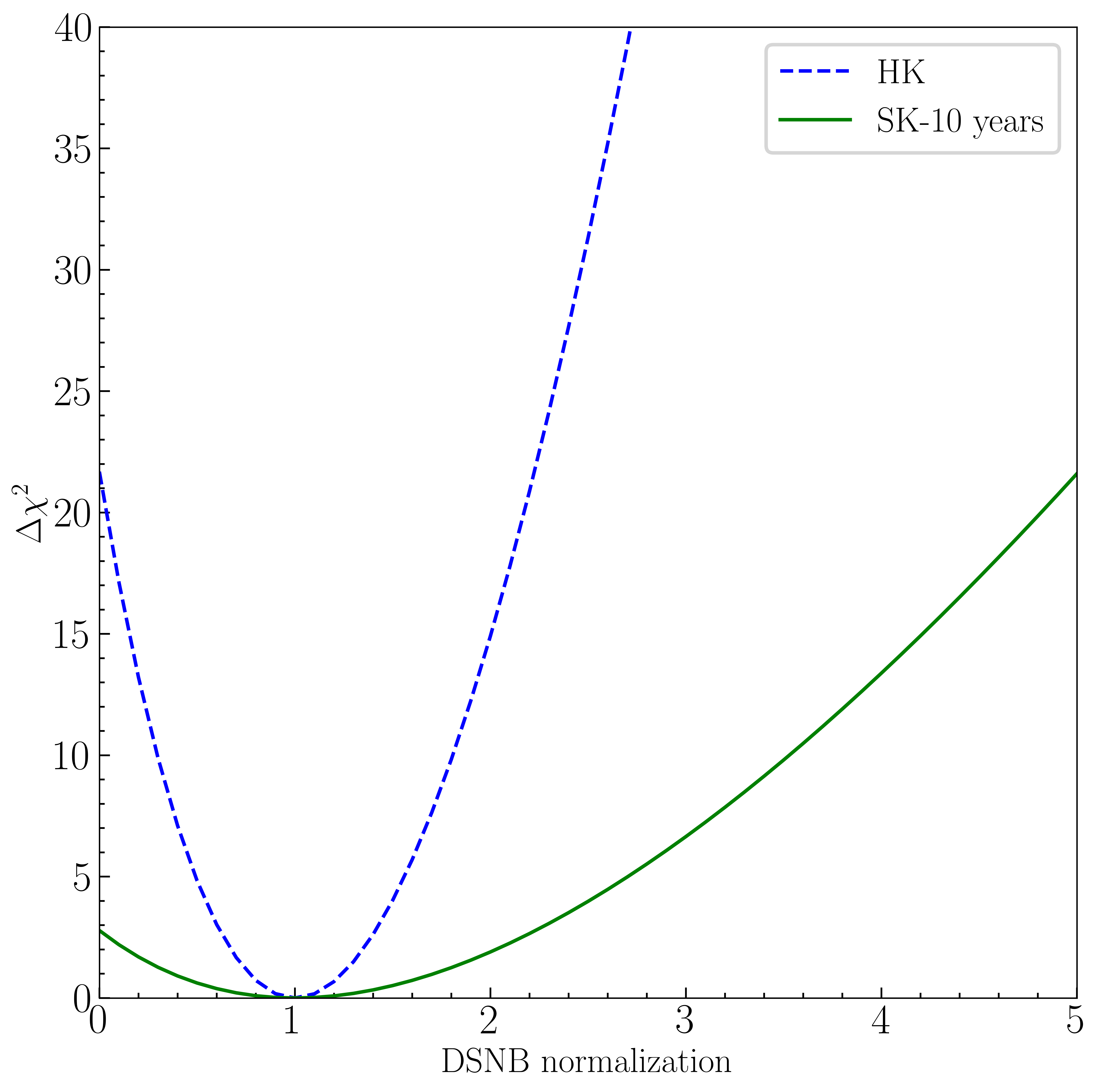}
    \caption{Projected sensitivity to the DSNB flux normalization for SK (10 years) and HK assuming ten years of data taking. See text for details. The DSNB normalization is defined relative to the Horiuchi et al.~(2009) model with an effective temperature of $6$~MeV~\cite{Horiuchi:2008jz}.}
    \label{fig:10_HK}
\end{figure}

We  use our framework to estimate the sensitivity of future experiments to the DSNB flux. We consider a Gd-loaded detector with the same specifications as SK-VII but an exposure of 10 years and consider 10 years of HK data, assuming similar neutron-tagging capabilities. To compute these sensitivities, we repeat the analysis of SK-VII changing the exposure accordingly and substitute the data with mock data computed assuming it is consistent with the Horiuchi et al.~(2009) model with an effective temperature of $6$~MeV~\cite{Horiuchi:2008jz} (defined as a DSNB normalization of 1). Our results are depicted in Figure \ref{fig:10_HK}. Assuming the best-fit point does not change, ten years of SK data should translate into a hint for the DSNB. (The zero flux scenario would be associated with a $\Delta\chi^2 \simeq 3.5$.) On the other hand, ten years of HK data are very sensitive to the DSNB. In particular, assuming the flux is consistent with the current best-fit obtained by SK-VI plus SK-VII, HK should exclude the no-DSNB hypothesis at almost the five-sigma level ($\Delta\chi^2 \simeq 22$).

\subsection{Ruling out the DSNB}\label{sec:alpha}

\begin{figure}
    \centering
    \includegraphics[width=\columnwidth]{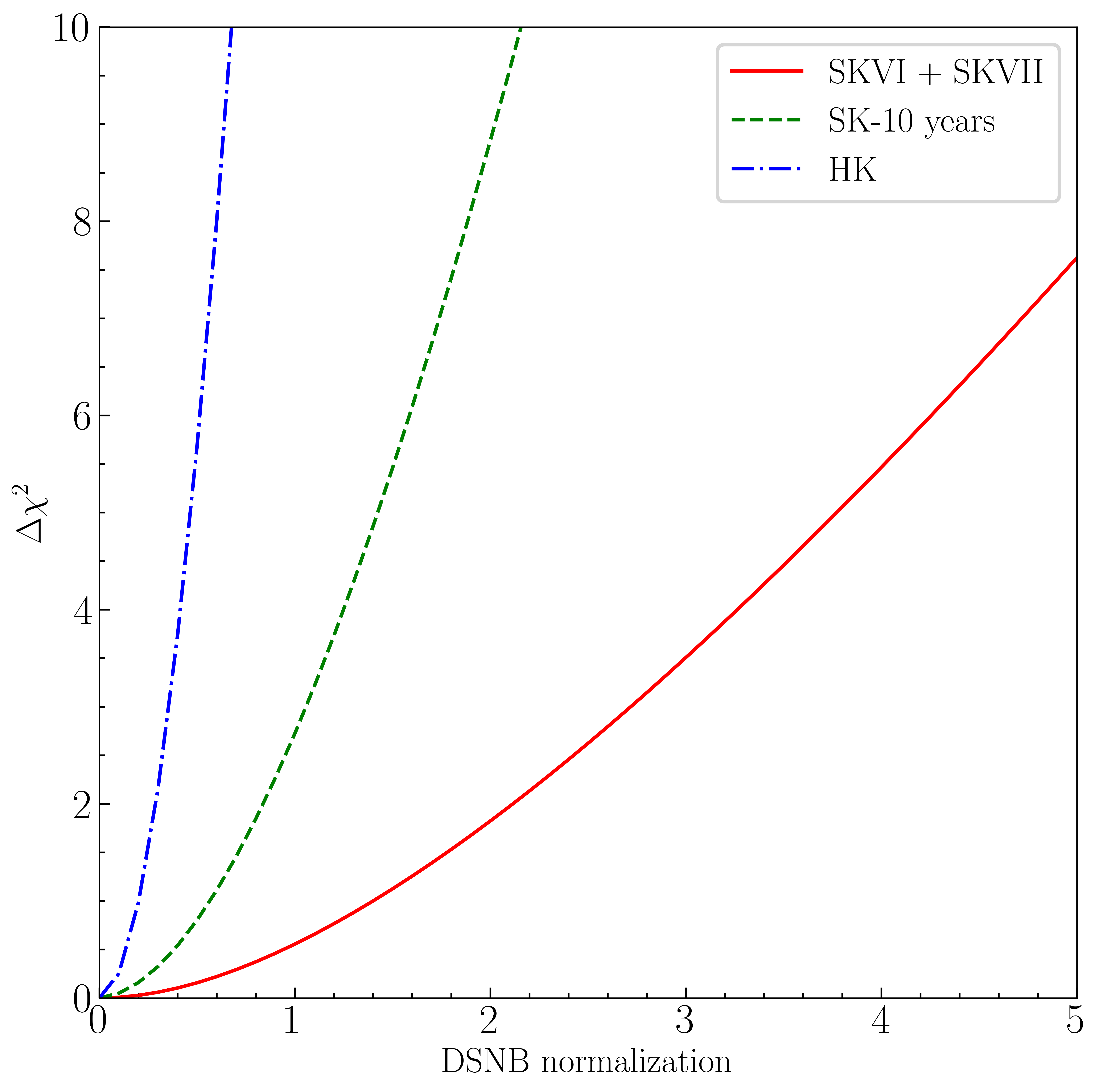}
    \caption{Sensitivity from SK-VI+SK-VII and future projections for SK(10 years) and HK to the DSNB normalization assuming no-DSNB hypothesis. The DSNB normalization is defined relative to the Horiuchi et al.~(2009) model with an effective temperature of $6$~MeV~\cite{Horiuchi:2008jz}.}
    \label{fig:NO}
\end{figure}

We also consider the hypothesis that the DSNB flux is, in fact, absent. We repeat our analyses for mock data consistent with no DSNB. Our results are depicted in Figure~\ref{fig:NO} for an exposure consistent with SK-VI plus SK-VII (solid, red line) ten years of SK (dashed, green line) and ten years of HK (dot-dashed, blue  line).

Keeping in mind that DSNB models are associated with DSNB normalizations of order a few, current data has limited power to disfavor our current understanding of the DSNB. Ten years of SK data would be able to modestly disfavor (around two sigma) a DSNB normalization of 1 and ten years of HK data would decisively rule out most models for the DSNB~\cite{Super-Kamiokande:2025sxh}.

\subsection{Measuring the Shape of the  DSNB Flux}

After the DSNB flux is observed, the next step is to start characterizing it. We explored the ability of SK-VI and SK-VII data, 10 years of SK data, and 10 years of HK data to measure the effective neutrino temperature $T_{\rm eff}$ assuming the DSNB flux, at the energies of interest, can be parameterized by an exponential, as discussed in Section~\ref{sec:dsnbflux}. In order to perform the fit, we modify Eq.~\eqref{expflux} as follows
\begin{equation}\label{expflux_mod}
    \Phi (E)=k\cdot\Phi_0\exp (-(E-E_0)/T_{\rm eff}),
\end{equation}
where $k$ is a dimensionless constant that allows us to control the normalization of the DSNB while keeping $\Phi_0$ and $E_0$ fixed. These are taken from Ref.~\cite{Horiuchi:2008jz}. In particular, we fix $(E_0,\Phi_0) = (25.49 \rm \, MeV, 0.049\, {\rm cm^{-2}s^{-1}MeV^{-1}})$. The results discussed here do not depend strongly on this choice.

\begin{figure}[h]
    \centering
    \includegraphics[width=\columnwidth]{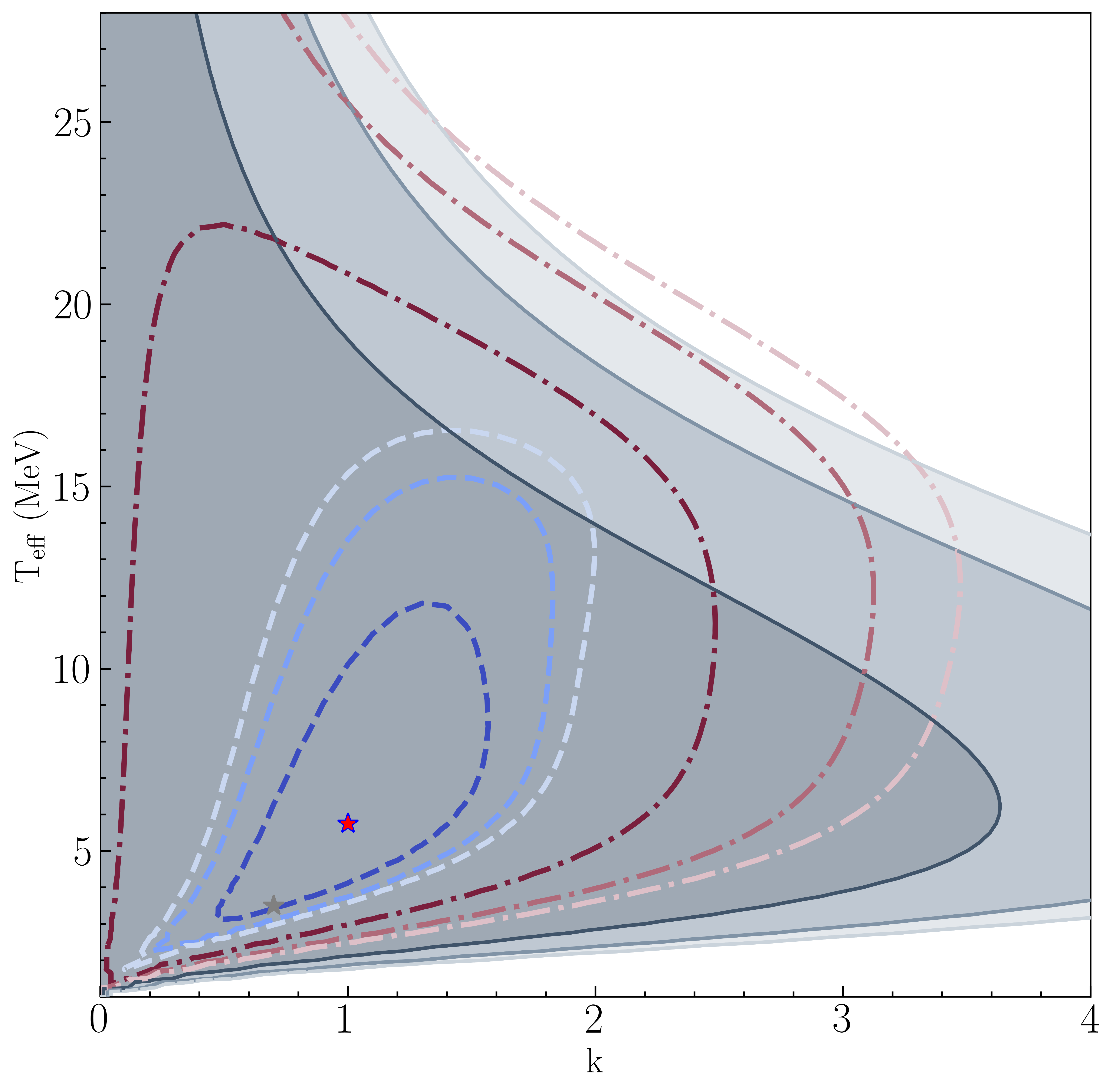}
    \caption{$1\sigma,90\%,\text{and }2\,\sigma$ regions (darkest, lighter, lightest colors, respectively) for 2 \textit{d.o.f.} for the $(k,T_{\rm eff})$ fit to SK-VI + SK-VII data (gray regions) and Asimov data for a 10-year SK projection (red dash-dot lines) and for HK (blue dashed lines). See text for details. The gray, red, and blue stars mark the best-fit points for SK-VI + SK-VII, SK (10 years), and HK (10 years), respectively. The best-fit points for HK (10 years) and SK (10 years) coincide because we are fitting the same model, namely the Horiuchi et al.~(2009) model with an effective temperature of $6$~MeV~\cite{Horiuchi:2008jz}.}
    \label{fig:alpha_exp_v2}
\end{figure}

We follow the prescription of Section~\ref{sec:results_overall}, using Asimov data generated with the 6 MeV, min.~flux from Ref.~\cite{Horiuchi:2008jz}, and simultaneously fit for $T_{\rm eff}$ and $k$. Figure \ref{fig:alpha_exp_v2} depicts the results of this fit for the combined SK-VI plus SK-VII data, ten years of SK data, and ten years of HK data. Ten years of HK data will be able to effectively constrain this parameter space. For HK, the effective temperature at the best-fit point is $T_{\rm eff}^{\rm best\,fit } = 5.75\,{\rm MeV}$ and can be compared to the theoretical expectation obtained by plugging into Eq.~\eqref{Teff} a neutrino temperature $T_\nu = 6\,{\rm MeV}$, an energy scale $E_0 = 20\, {\rm MeV}$, and an effective redshift $z_{\rm eff}=1$. In this case, $T_{\rm eff}^{\rm expected}=4.29\,{\rm MeV}$, well within the $1\,\sigma$ region allowed by HK. 

\subsection{Impact of Oscillations}

Flavor oscillations and other aspects of neutrino transport impact the atmospheric neutrino background and, consequently, the sensitivity of SK and HK to the DSNB flux. Here, we explore this relationship in more detail. 

First, we estimate how the sensitivity to the DSNB depends on the details of neutrino flavor oscillations. For that, we compare the standard scenario, in which flavor oscillations are computed numerically using the PREM model for the Earth and including neutrino propagation through the atmosphere, with a simplified scenario that, while less realistic, allows for faster calculations. 

In the simplified scenario, we assume that the oscillation probabilities ``average out'' so that they do not depend on the neutrino energy or the baseline (i.e., the neutrino direction). A natural choice is to use the input from averaged-out vacuum oscillations: $P_{\alpha\beta} = \sum_{i} |U_{\alpha i}|^2 |U_{\beta i}|^2$, $\alpha,\beta=e,\mu,\tau$. In this case, the event rate from atmospheric neutrinos is suppressed by approximately $8\%$ over the entire energy range. This effect can be partially compensated by the normalization uncertainty of the flux. 

\begin{figure}
    \centering
    \includegraphics[width=\columnwidth]{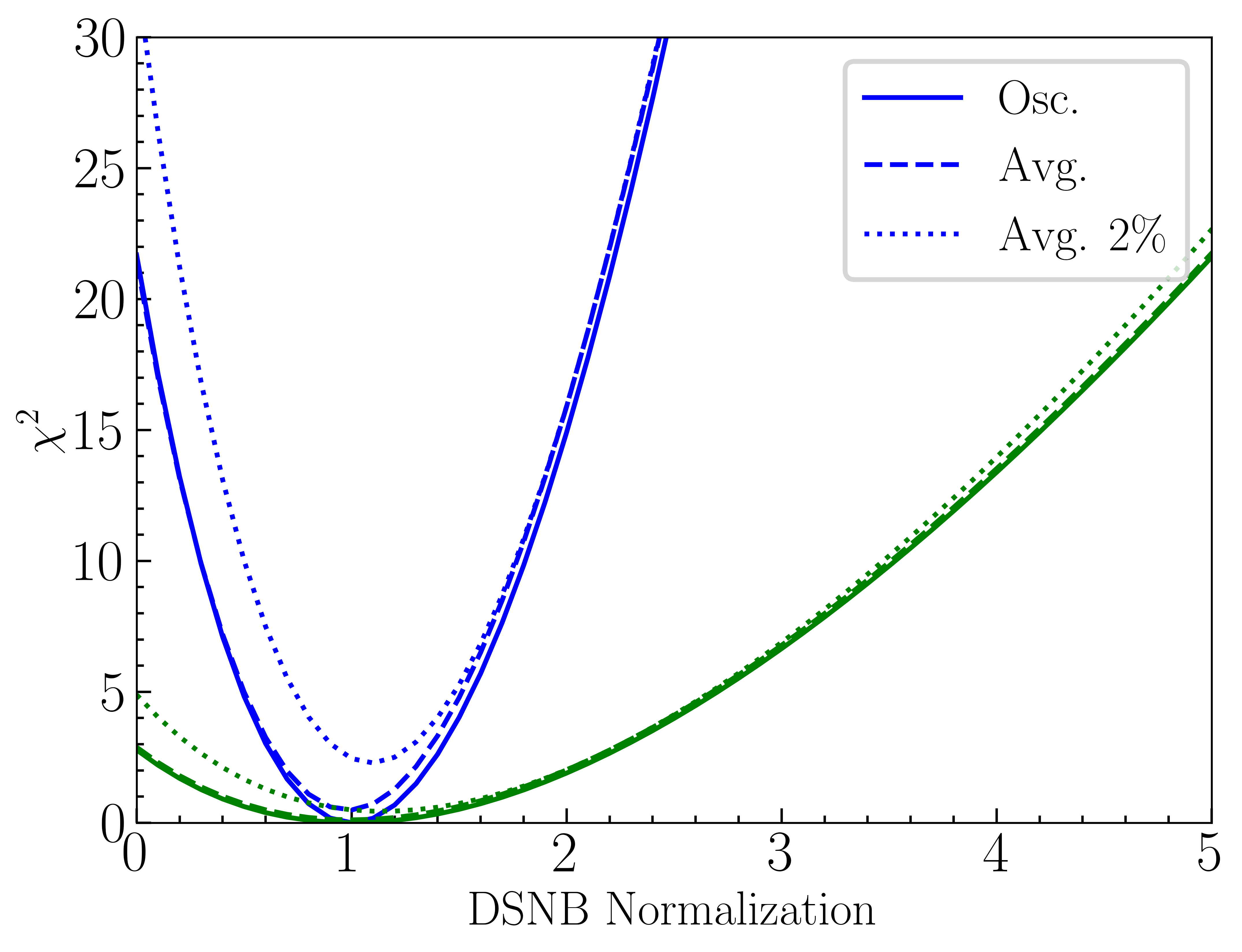}
    \caption{Sensitivity to the DSNB assuming 10 years of SK (green) and HK (blue). The DSNB normalization is defined relative to the Horiuchi et al.~(2009) model with an effective temperature of $6$~MeV~\cite{Horiuchi:2008jz}. We compare the sensitivity for three scenarios: full oscillation treatment (solid), averaged vacuum oscillations (dashed), and averaged oscillations assuming an improved  $2\%$ normalization uncertainty on the atmospheric neutrino flux (dotted).}
    \label{fig:oscavg}
\end{figure}

Fig.~\ref{fig:oscavg} depicts the sensitivity to the DSNB if one analyses the simulated data, consistent with standard oscillations, assuming averaged-out flavor oscillations, along with the results of the standard analysis (Fig.~\ref{fig:10_HK}). For the simplified analysis, the global minimum slightly increases and the sensitivity is slightly decreased. Since averaged-out oscillations are partially degenerate with an overall normalization shift, we also consider the case in which the atmospheric flux uncertainty is reduced to $2\%$ (`Avg.~2\%' in Fig.~\ref{fig:oscavg}). In this scenario, we find a preference for larger values of the DSNB normalization.

Since oscillations modify the atmospheric neutrino background in a meaningful way, it is interesting to investigate whether the low-energy IBD-like events collected for the DSNB search can be used to ``see'' neutrino oscillations and measure oscillation parameters. Given that, here, we integrate over all directions and matter effects are subdominant, we expect negligible sensitivity to the mass-squared splittings but expect some sensitivity to $\sin^2\theta_{23}$, which primarily affects the normalization of the different atmospheric background components, as discussed in previous studies~\cite{Peres:2009xe}. In Appendix~\ref{mixing}, we present the results of such an analysis. They reveal that SK-VI plus SK-VII and future data have limited sensitivity to $\sin^2\theta_{23}$. 

\begin{figure}
    \centering
    \includegraphics[width=\columnwidth]{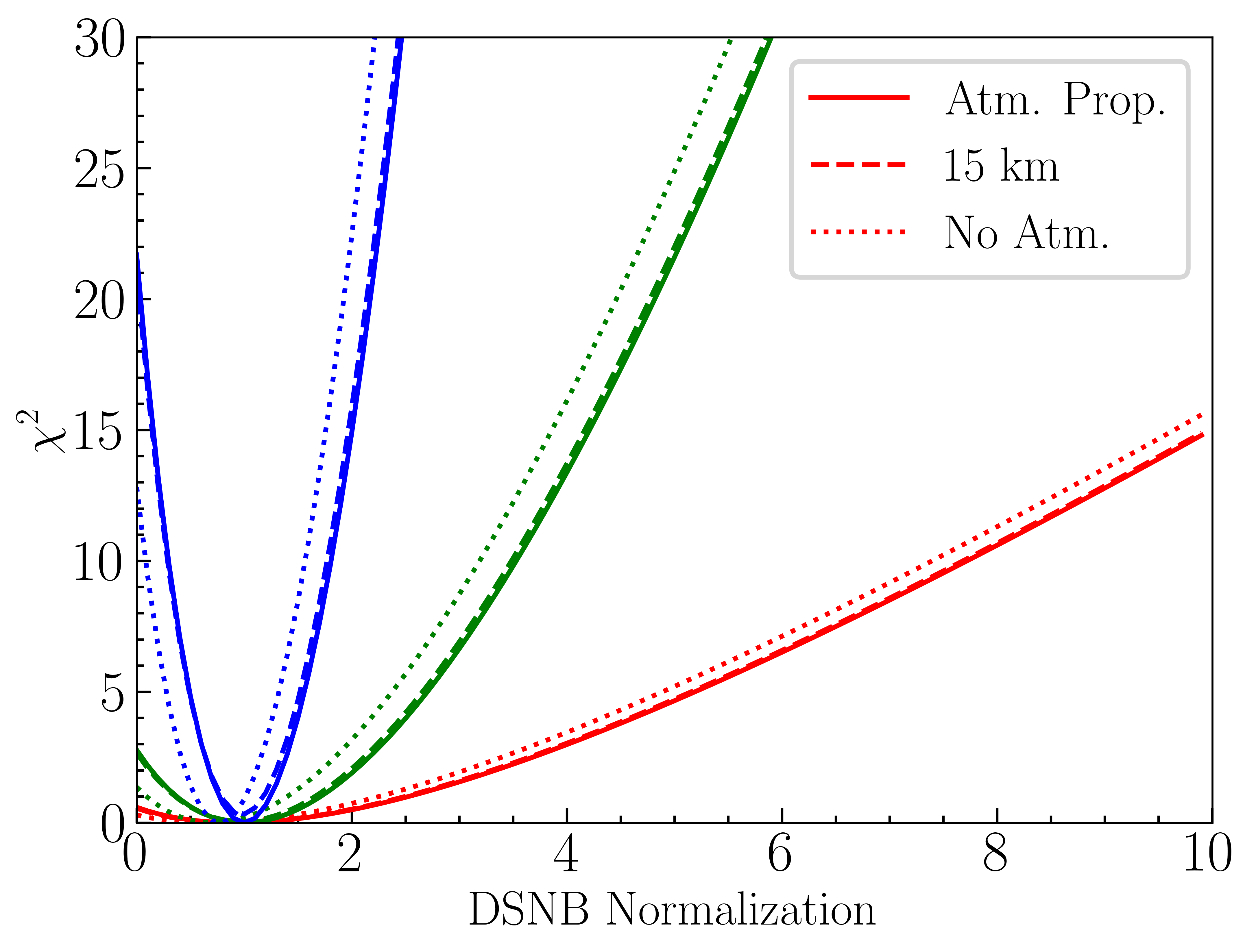}
    \caption{Sensitivity to the DSNB assuming current SK data (red), 10 years of SK (green) and HK (blue). The DSNB normalization is defined relative to the Horiuchi et al.~(2009) model with an effective temperature of $6$~MeV~\cite{Horiuchi:2008jz}. We consider three scenarios: one in which neutrinos are produced at different altitudes (Atm.\ Prop.), one in which all neutrinos are assumed to be produced at a fixed altitude of 15~km, and one in which neutrino propagation through the atmosphere is neglected.}
    \label{fig:oscatm}
\end{figure}

Finally, since roughly half of the atmospheric neutrinos originate from down-going directions, we also investigate the impact of their propagation through the atmosphere. To this end, we consider two simplified scenarios: one in which all neutrinos are assumed to be produced at a fixed altitude of 15~km, and another in which the production altitude is set to zero. The results obtained when making these simplifying assumptions are depicted in Fig.~\ref{fig:oscatm}, along with the results of the standard analysis (Fig.~\ref{fig:10_HK}), where we allow for neutrinos to be produced at different heights. 

Assuming a fixed neutrino production altitude of 15~km does not significantly impact the current results or the projected sensitivity\footnote{Note that we adopt a single model for the atmospheric density distribution~\cite{USSA1976}. This conclusion may change when considering alternative models.}. In contrast, neglecting propagation through the atmosphere (production altitude equal to zero) leads to an enhancement of the atmospheric background since $P_{\mu\mu}=1$ for down-going neutrinos and ``shifts'' the fit towards lower DSNB fluxes, translating into an incorrect estimate of the DSNB flux. Furthermore, in this simplified analysis, the sensitivity to exclude the no-DSNB hypothesis is reduced by nearly a factor of two. The impact of neglecting the atmosphere in the analysis is more severe with increased statistics.

\section{Conclusions}
\label{sec:conclusions}

The observation of the DSNB would open a new window for particle astrophysics. While current data published by SK do not show statistically significant evidence for the existence of the DSNB, expectations are that, with increased statistics, it will eventually be detected.

In this work, we reproduced the analysis carried out and made available by SK using data from phases VI and VII \cite{Super-Kamiokande:2025sxh}. Building on these results, we estimated SK's sensitivity to the DSNB assuming it collects 10 years of data consistent with those collected in phase VII. Looking further into the future, we also estimate the sensitivity of ten years of HK data. Our results indicate that a 10-year exposure at SK would provide at most a relatively weak preference for the existence of the DSNB. A robust detection and precise characterization of the DSNB flux will require very large, next-generation detectors like HK, assuming it has neutron-tagging capabilities similar to those of SK loaded with gadolinium.

We also studied the impact of atmospheric neutrino flavor oscillations on the measurement of the DSNB. We find that these effects become increasingly important as experimental precision improves. In particular, an inaccurate modeling of neutrino propagation through the atmosphere, especially at energies of tens of MeV, can lead to non-negligible biases in the inferred DSNB flux.

\section{Acknowledgments}
We would like to thank Sergio Palomares-Ruiz for useful discussions. PBM would like to thank IPPP and CERN for hospitality and support during his visits, where part of this work was carried out. The work of PBM is supported by the Spanish MIU through the National Program FPU (grant number FPU22/03600). This work has been partially supported by the Spanish Research Agency through grant CNS2023-145338 funded by MCIN/AEI/10.13039/501100011033 and by “European Union NextGenerationEU/PRTR”, through Grants PID2022-142545NB-C21 and PID2025-171019NB-C21  funded by MCIN/AEI/10.13039/501100011033/ FEDER, UE, and through the Grant IFT Centro de Excelencia Severo Ochoa No. CEX2020-001007-S, funded by MCIN/AEI/10.13039/501100011033. This work is partially funded by the European Commission – NextGenerationEU, through Momentum CSIC Programme: Develop Your Digital Talent. Numerical calculations have been performed on the Hydra cluster at IFT. We acknowledge HPC support by Emilio Ambite, staff hired under the Generation D initiative, promoted by Red.es, an organization attached to the Spanish Ministry for Digital Transformation and the Civil Service, for the attraction and retention of talent through grants and training contracts, financed by the Recovery, Transformation and Resilience Plan through the European Union’s Next Generation funds. IMS was supported by grant 1001110342 funded by MICIU/AEI/10.13039/501100011033 and by ESF+ and by grant PID2025-172338NB-I00 funded by MICIU/AEI/ 10.13039/501100011033 and by ERDF,EU. AdG was supported in part by the US Department of Energy grant \#de-sc0010143 and in part by the National Science Foundation grant PHY-1630782. AdG also acknowledges the Center for Theoretical Underground Physics and Related Areas (CETUP), The Institute for Underground Science at Sanford Underground Research Facility (SURF), and the South Dakota Science and Technology Authority for hospitality and financial support while some of this work was carried out.


\appendix
\section{Validation of neutrino fluxes}
\label{sec:app-flux}

Here we provide more details of our estimate of the atmospheric neutrino flux, along with a comparison to well-established results. 
 
\begin{figure}
    \centering
    \includegraphics[width=\columnwidth]{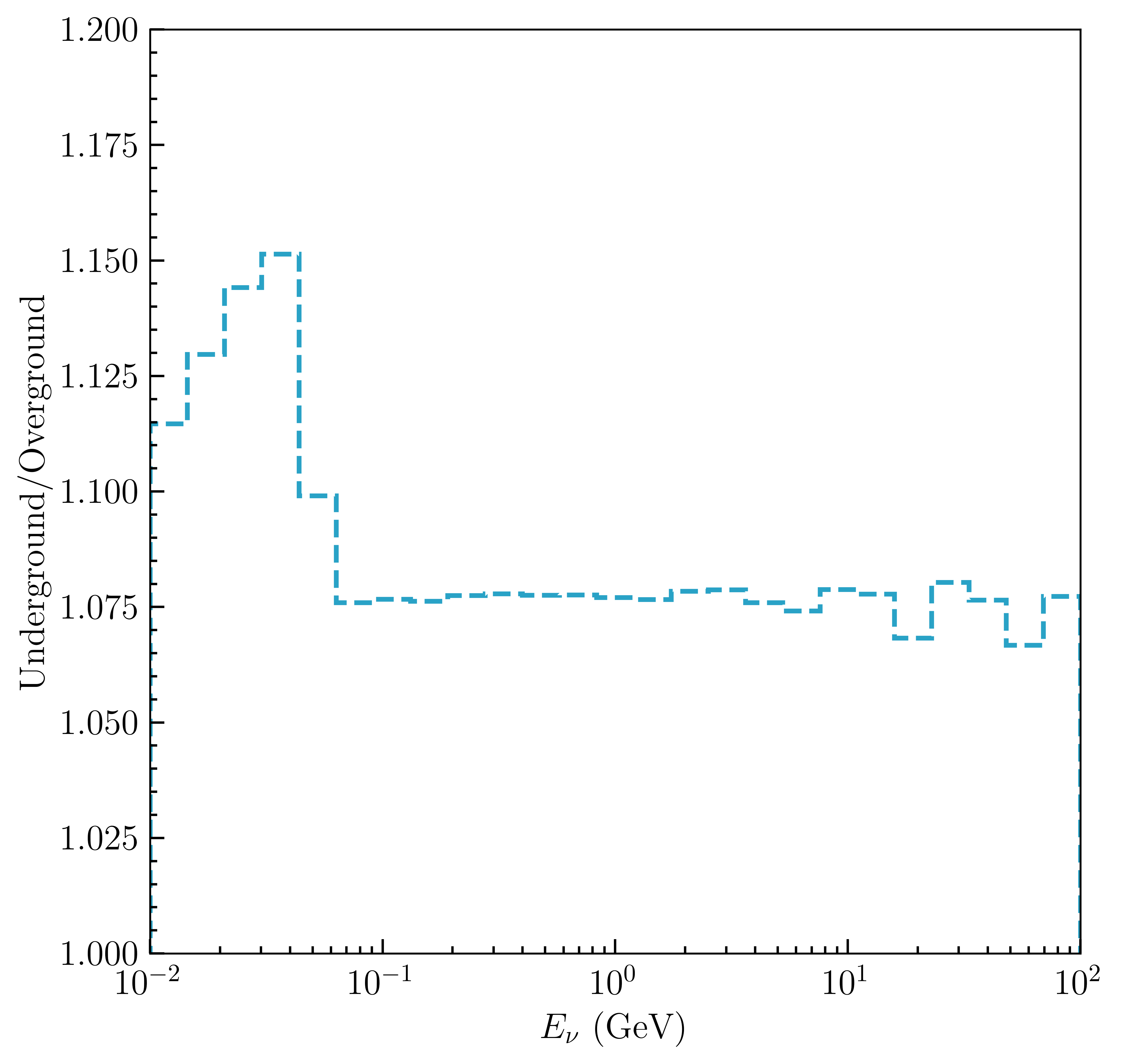}
    \caption{Ratio of the expected $\nu_e$ fluxes at ground level and underground, without including neutrino oscillation effects.}
    \label{fig:nue_ratio}
\end{figure}

 Figure~\ref{fig:nue_ratio} depicts the ratio of the $\nu_e$ flux underground relative to the flux at the surface. The silicon layer allows the remaining mesons to decay, increasing the number of measured neutrinos. Furthermore, for neutrino energies around 30~MeV, there is a ``bump'' in the flux of underground neutrinos caused by high-energy muons reaching the ground and decaying at rest.

\begin{figure*}
    \centering
    \includegraphics[width=\textwidth]{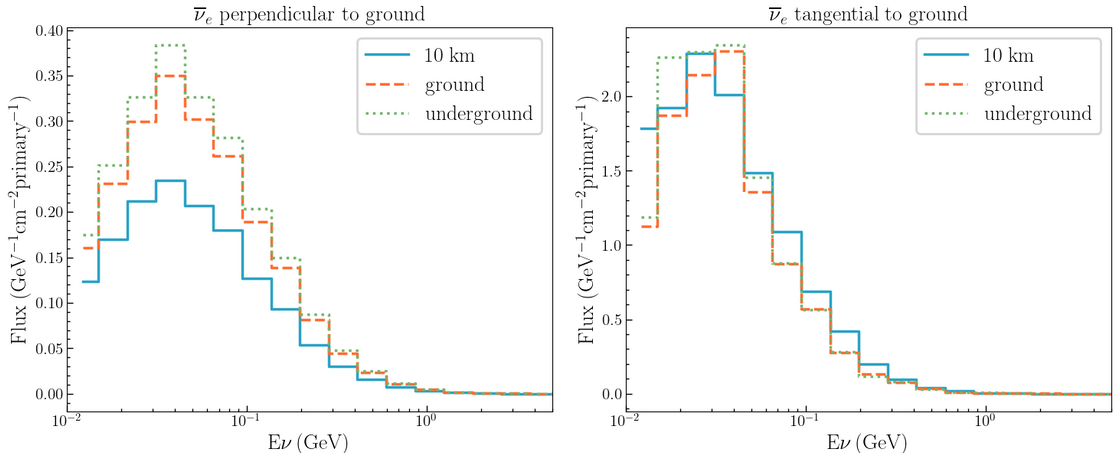}
    \caption{Comparison of electron neutrino fluxes at various heights incident perpendicular (left) or tangential (right) to the detector's surface obtained with our Fluka simulation.}
    \label{fig:angsol}
\end{figure*}

Figure~\ref{fig:angsol} depicts the flux of neutrinos that arrive perpendicular and tangential to the Earth's surface. For some neutrino energies, there are more tangential neutrinos underground than on the surface. The neutrinos from the surface aren't ``missing," they just entered the surface with a direction `too parallel'' to the ground and escape the detector volume. 

\begin{figure}
\centering
    \includegraphics[width=\columnwidth]{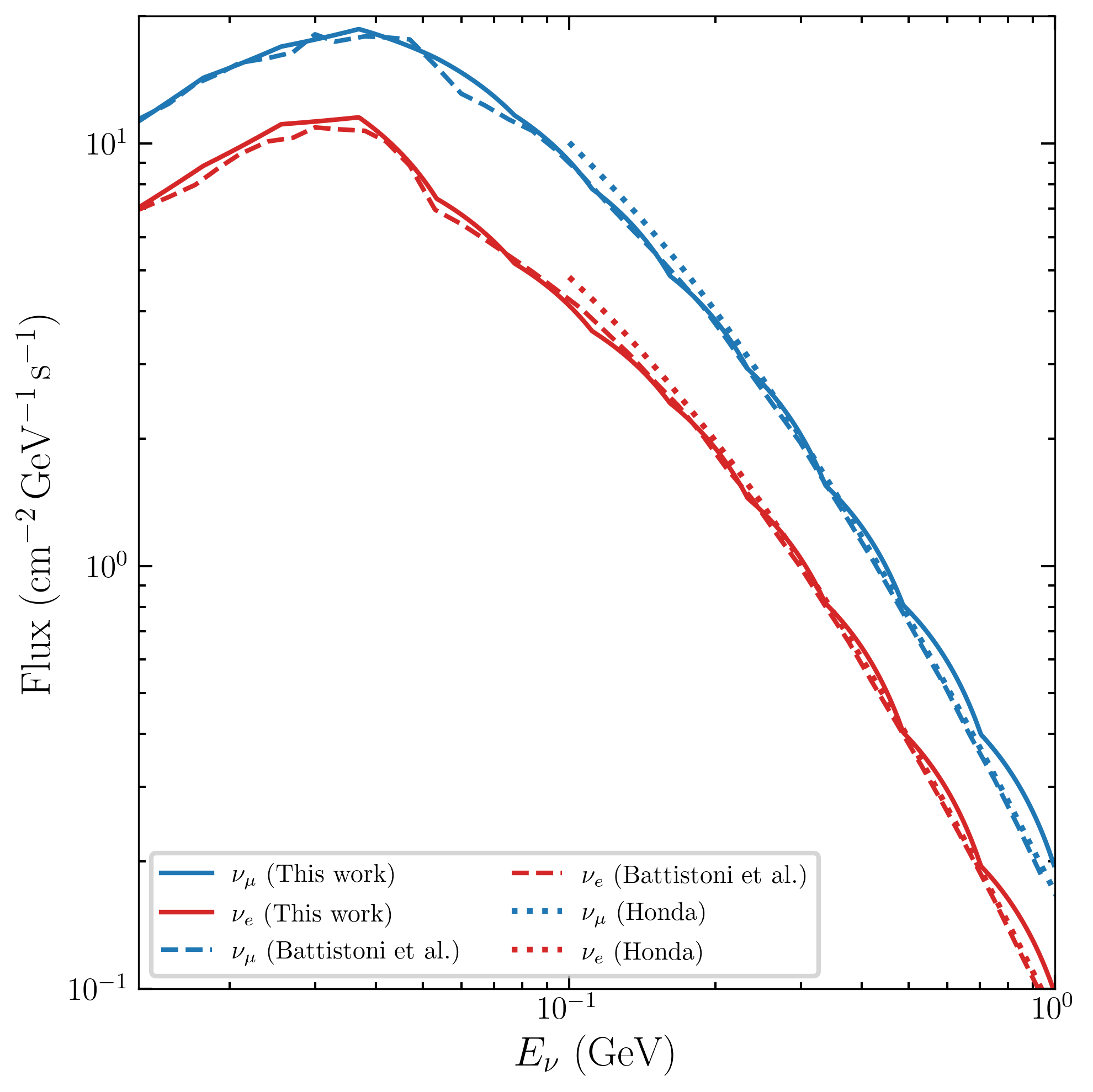}
    \caption{Comparison between our calculated underground neutrino fluxes and literature benchmarks (Honda~\cite{PhysRevD.92.023004} and Battistoni et~al.~\cite{BATTISTONI2005526}). Blue (red) curves denote $\nu_\mu$ ($\nu_e$) fluxes. Solid lines correspond to this work, dashed lines to Battistoni et~al., and dotted lines to Honda. Antineutrino channels show an equivalent level of agreement and are omitted for visual clarity.}
\label{fig:flux_comparison}
\end{figure}

Finally, we compare our estimates for the underground neutrino fluxes with existing calculations in the literature, specifically the Honda flux~\cite{PhysRevD.92.023004} (table 1.1b, available \href{http://www-rccn.icrr.u-tokyo.ac.jp/mhonda/public/nflx2014/index.html}{here}) and the FLUKA simulations by Battistoni et~al.~\cite{BATTISTONI2005526}. For visual clarity, Fig.~\ref{fig:flux_comparison} displays this comparison only for the representative $\nu_\mu$ and $\nu_e$ channels. The corresponding antineutrino fluxes ($\bar{\nu}_\mu$ and $\bar{\nu}_e$) were also verified and exhibit an equally good level of agreement. Overall, the match across all flavors is more than satisfactory for our purposes.

\section{Oscillation probabilities}
\label{sec:app-osc}

\begin{figure*}
    \centering
    \includegraphics[width=\columnwidth]{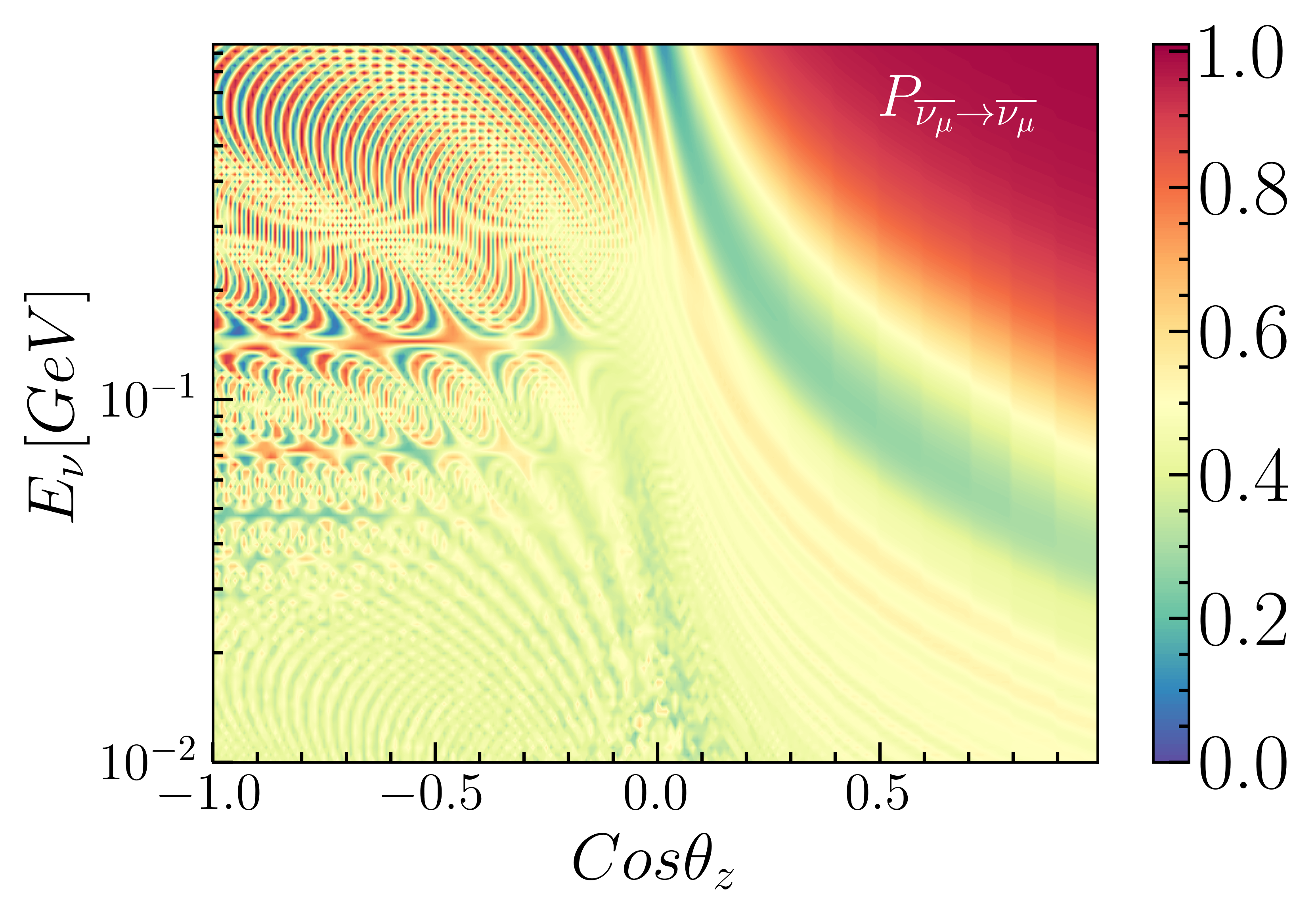}
    \includegraphics[width=\columnwidth]{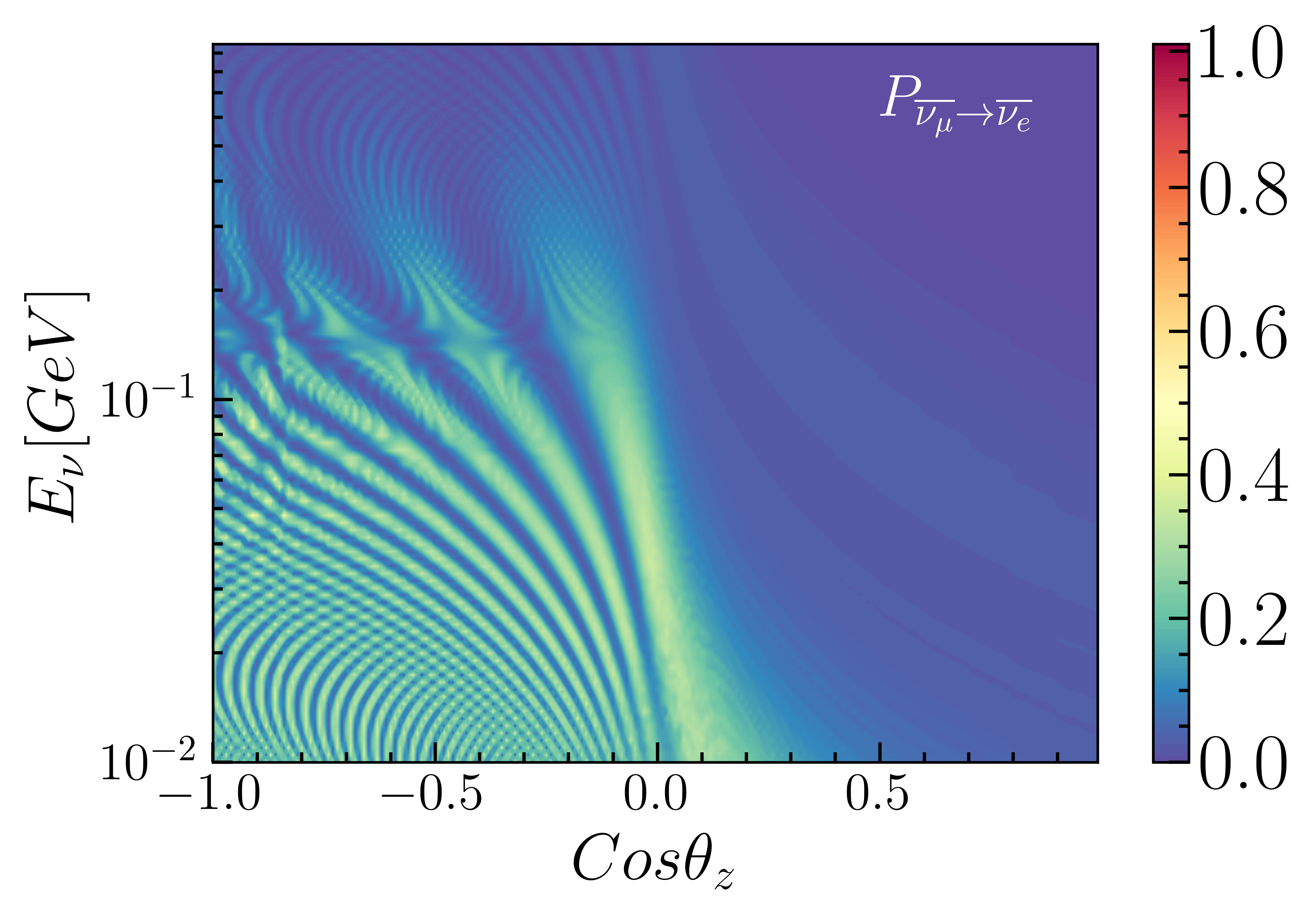}
    \includegraphics[width=\columnwidth]{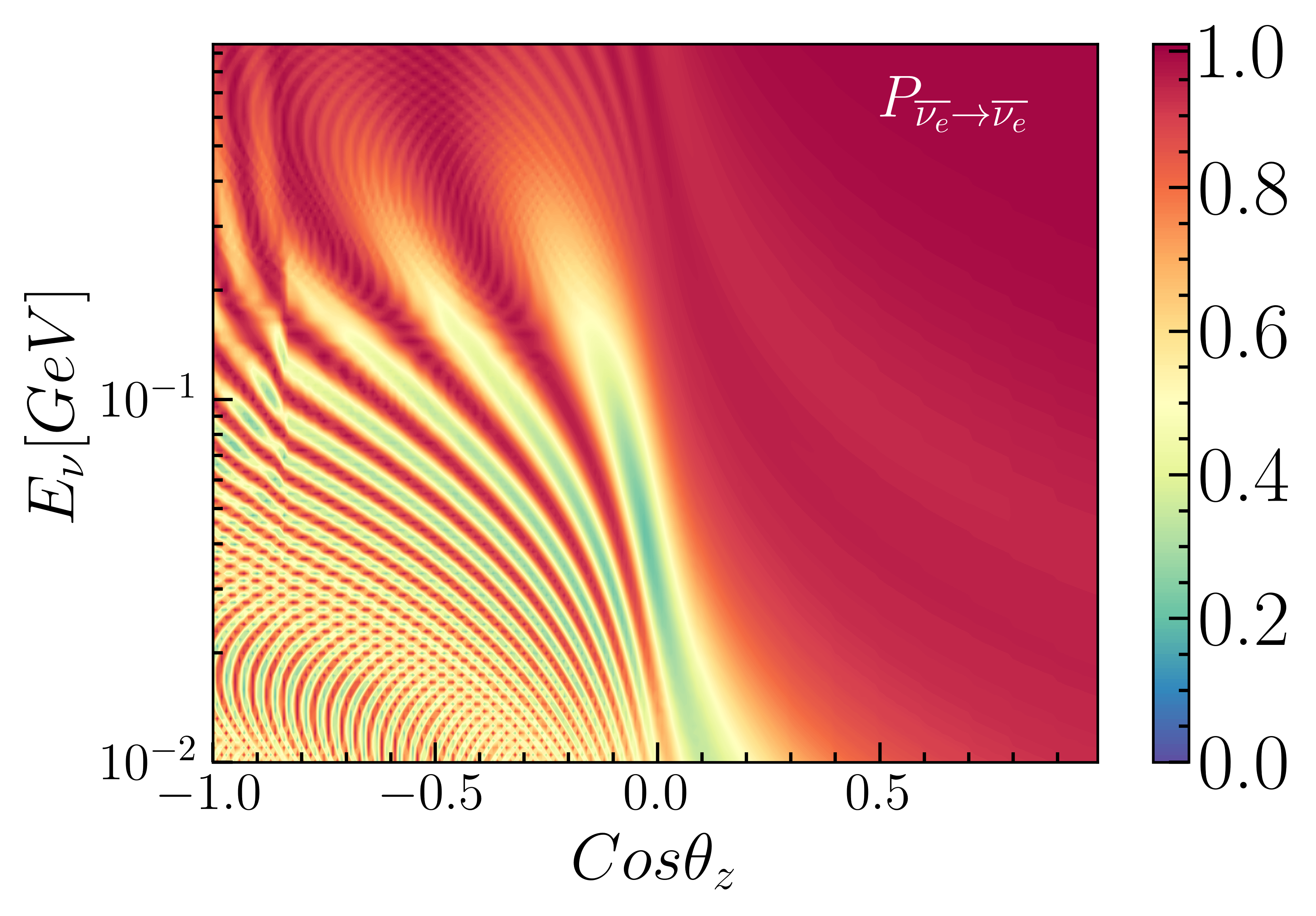}
    \caption{Oscillograms for sub-GeV antineutrinos including the electron-antineutrino survival probabilities, the muon antineutrino survival probabilities, and the muon antineutrino to electron antineutrino appearance probabilities.}
    \label{fig:Osc2}
\end{figure*}

The propagation of atmospheric neutrinos through the Earth and the atmosphere is impacted by flavor oscillations. We study these oscillations for energies below the GeV scale and for all baselines, ranging from up-going to down-going neutrinos. Fig.~\ref{fig:Osc2} depicts oscillograms for the three channels most relevant for the determination of the DSNB background: the muon antineutrino survival probability ($P_{\bar{\nu}_{\mu}\rightarrow\bar{\nu}_{\mu}}$), the electron antineutrino survival probability ($P_{\bar{\nu}_{e}\rightarrow\bar{\nu}_{e}}$), and the electron antineutrino appearance probability ($P_{\bar{\nu}_{\mu}\rightarrow\bar{\nu}_{e}}$). For atmospheric propagation, we assume that neutrinos are produced at different altitudes and propagate through the atmosphere accordingly. We use the best-fit values for the oscillation parameters from NuFit~\cite{Esteban:2024eli} and assume the neutrino mass ordering is normal.

\section{SK IV + V + VI + VII + VIII} \label{combined}

Super-Kamiokande presented an update to their DSNB search at the Neutrino 2026 Conference, not yet published. It is a combined analysis of phases SK-IV, SK-V, SK-VI, SK-VII, SK-VII.5 and SK-VIII, corresponding to a total live-time of $5000$~days. The first two phases, which add up to  almost 70\% of the total live-time, correspond to data collected with pure water. SK-VI contains a $0.01\%$ Gadolinium concentration, and for the later phases the Gd concentration was increased to $0.03\%$.
While the detector efficiency changed among the different phases, we developed an effective simulation that fits the data presented at the conference using the detector efficiency of SK-VII. We present more details and our result here.  

In their new analysis, SK chose a lower energy threshold for the $n=1$ sample ($12$~MeV) and a higher energy threshold for the $n$-neutron sample ($22$~MeV). Due to the lower threshold, in the $n=1$ case, two additional backgrounds, due to spallation and neutral-current interactions, need to be introduced. On the other hand, the impact of these backgrounds is suppressed in the $n$-neutron sample. In our simulation, we took these backgrounds from the predictions provided by the collaboration and assigned a $10\%$ uncertainty to their normalization. We handled the atmospheric background in the same way we handled it in the SK-VI and SK-VII simulations, accounting for the correlation between the four atmospheric neutrino flux components reaching the detector, as predicted by the flux simulation, and the flavor oscillation probabilities. At the same time, we reduced the uncertainty on the atmospheric neutrino normalization to $10\%$.

\begin{figure}
    \centering
    \includegraphics[width=\columnwidth]{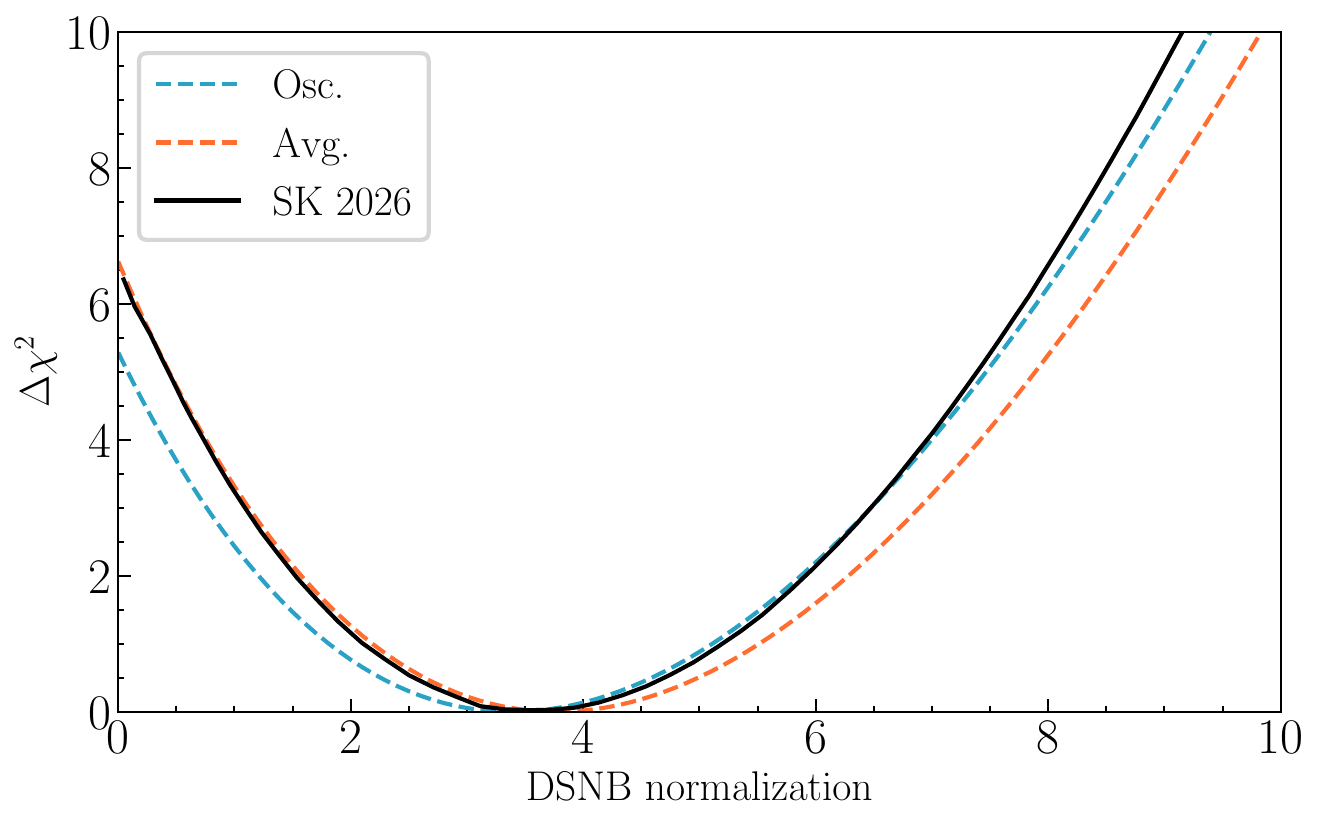}
    \caption{Combined analysis of the SK-IV, V, VI, VII, VII.5, and VIII phases. The DSNB normalization is defined relative to the Horiuchi et al.~(2009) model with an effective temperature of $6$~MeV~\cite{Horiuchi:2008jz}.  The result presented by the Super-Kamiokande Collaboration at the Neutrino 2026 Conference (solid, black, SK 2026) is compared to our estimates (dashed) corresponding to two different hypotheses about neutrino flavor evolution: averaged-out oscillations (red, Avg.) and a full treatment of oscillations along the neutrino trajectory (blue, Osc.).}
    \label{fig:neutrino}
\end{figure}

The results of this effective simulation, compared with those reported by Super-Kamiokande, are depicted in Fig.~\ref{fig:neutrino}. The DSNB normalization is defined relative to the Horiuchi et al.~(2009) model with an effective temperature of $6$~MeV~\cite{Horiuchi:2008jz}. We perform fits to the data under two hypotheses: averaged-out oscillations (Avg) and the full oscillation probability along the neutrino trajectory (Osc). Both hypotheses lead to similar results. We find that the null hypothesis can be excluded at $\Delta\chi^2 \sim 5$ when a full oscillation treatment is included; for averaged-out oscillations, the exclusion is slightly stronger, by about $1.5$ units of $\chi^2$. We compare our results with those presented by the Super-Kamiokande Collaboration at the Neutrino 2026 Conference (SK 2026) and find good agreement. 

\section{Sensitivity to $\sin^2\theta_{23}$} \label{mixing}

The impact of neutrino oscillations on the sensitivity to the DSNB, discussed earlier, invites us to investigate whether it is possible to constrain neutrino oscillation parameters using the DSNB-related event sample. Given the absence of directional information, we concentrate our discussion on the sensitivity to  the mixing parameter $\sin^2\theta_{23}$, which primarily affects the normalization of the different atmospheric background components. These effects were discussed previously in the literature~\cite{Peres:2009xe}.

\begin{figure}
    \centering
    \includegraphics[width=\columnwidth]{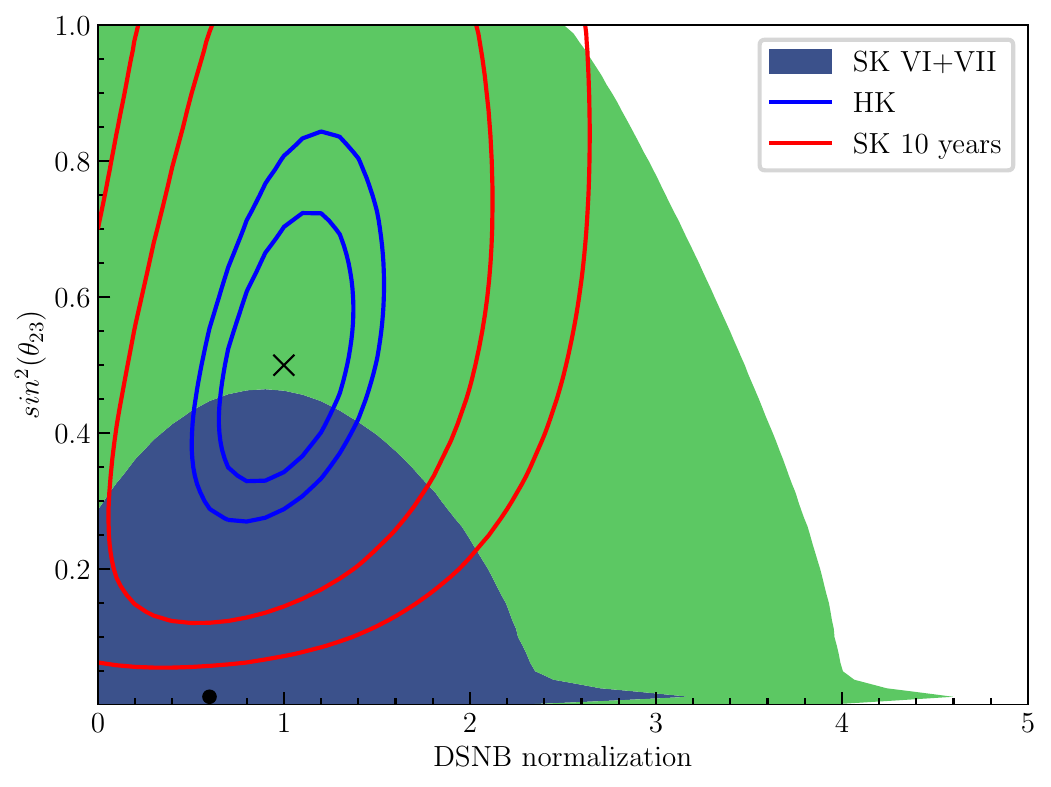}
    \caption{Joint constraints on $\sin^2\theta_{23}$ and the DSNB normalization from SK-VI and SK-VII data (colored regions), SK 10-year data (red lines), and HK (blue lines). The Different sized-regions correspond to the $1\sigma$ and 90\% confidence levels. The DSNB normalization is defined relative to the Horiuchi et al.~(2009) model with an effective temperature of $6$~MeV~\cite{Horiuchi:2008jz}.}
    \label{fig:osct23}
\end{figure}

We perform a joint fit for $\sin^2\theta_{23}$ and the normalization of the DSNB flux. All other oscillation parameters are held fixed. As everywhere else in this manuscript, the DSNB normalization is defined relative to the Horiuchi et al.~(2009) 
model with an effective temperature of $6$~MeV~\cite{Horiuchi:2008jz}. Our results are depicted in Fig.~\ref{fig:osct23} for the current SK data (SK-VI and SK-VII), SK-10 years (red lines),
and HK (blue lines). The current SK data exhibit a preference for small values of $\sin^2\theta_{23}$, driven by a preference of the data for a suppressed $\nu_e$ contribution. This effect may be attributed either to a statistical fluctuation or to an additional suppression at high energies resulting from the detector efficiency. Nevertheless, the entire range of $\sin^2\theta_{23}$ remains allowed at the 90\% confidence level. With 10 years of data and the SK-VII detector efficiency, very small values of $\sin^2\theta_{23}$ can be excluded. A substantially larger data set is required to exclude $\sin^2\theta_{23}=0$ or $1$  at more than the 90\% confidence level.
A more meaningful improvement requires a better understanding of the atmospheric neutrino flux and a reduction of its uncertainties.

The muon disappearance probability depends on $\sin^2 2\theta_{23}$, so any deviation from maximal mixing leads to an enhancement of the muon neutrino flux. This octant degeneracy is broken by the electron neutrino contribution, which depends on $\sin^2\theta_{23}$ and leads to a suppression of events in the lower octant and an enhancement in the higher octant. This explains why the sensitivity to $\sin^2\theta_{23}$ is asymmetric: Very small values of $\sin^2\theta_{23}$ correspond to an increase in muon events and a reduction in electron events, which is more difficult to accommodate within the systematic uncertainties given the assumed $2\%$ uncertainty in the flavor ratio. For values of $\sin^2\theta_{23}$ in the higher octant, both the electron and muon contributions are enhanced, and this effect can be compensated by adjusting the overall flux normalization.

\bibliography{References}

\end{document}